\documentclass[11pt]{article}
\usepackage[english]{babel}
\usepackage{amssymb,amsmath,graphicx}
\usepackage{amsthm}
\usepackage{float}
\usepackage{afterpage}
\usepackage{color}
\usepackage[round,authoryear]{natbib}
\usepackage{enumitem}
\usepackage{setspace}
\usepackage{algorithmic}
\usepackage{algorithm}
\usepackage[table]{xcolor}
\usepackage{multirow}
\usepackage{natbib}
\usepackage{microtype}
\newcolumntype{M}[1]{>{\centering}m{#1}}
\newcolumntype{H}{>{\setbox0=\hbox\bgroup}c<{\egroup}@{}}

 \bibpunct[, ]{(}{)}{,}{a}{}{,}%
 \def\newblock{\ }%

\newcolumntype{d}[1]{D{.}{.}{#1}}

\newcommand{\cV}{{\mathcal V}}

\newcommand{\cO}{{\mathcal O}}

\newcommand{\cS}{{\mathcal S}}

\newcommand{\cI}{{\mathcal I}}

\title{A study on exponential-size neighborhoods for the bin packing problem with conflicts}

\author{Renatha Capua, Yuri Frota, Luiz Satoru Ochi, Thibaut Vidal}

\begin{document}

\begin{center}

\vspace*{-1.2cm}

\begin{LARGE}
A study on exponential-size neighborhoods for \vspace*{0.2cm}

the bin packing problem with conflicts
\end{LARGE}

\vspace*{0.4cm}

\textbf{Renatha Capua} \\
Instituto de Computa\c{c}\~ao -- Universidade Federal Fluminense \\
Rua Passo da P\'{a}tria, 156 - S\~{a}o Domingos, Niter\'{o}i - RJ, 24210-240, Brazil \\
rcapua@ic.uff.br \\
\vspace*{0.2cm}
\textbf{Yuri Frota} \\
Instituto de Computa\c{c}\~ao -- Universidade Federal Fluminense \\
Rua Passo da P\'{a}tria, 156 - S\~{a}o Domingos, Niter\'{o}i - RJ, 24210-240, Brazil \\
yuri@ic.uff.br \\
\vspace*{0.2cm}
\textbf{Luiz Satoru Ochi} \\
Instituto de Computa\c{c}\~ao -- Universidade Federal Fluminense\\
Rua Passo da P\'{a}tria, 156 - S\~{a}o Domingos, Niter\'{o}i - RJ, 24210-240, Brazil \\
satoru@ic.uff.br \\
\vspace*{0.2cm}
\textbf{Thibaut Vidal} \\
Departamento de Inform\'{a}tica, Pontif\'{i}cia Universidade Cat\'{o}lica do Rio de Janeiro (PUC-Rio) \\
Rua Marqu\^{e}s de S\~{a}o Vicente, 225 - G\'{a}vea, Rio de Janeiro - RJ, 22451-900, Brazil \\
vidalt@inf.puc-rio.br \\

\vspace*{0.5cm}

\begin{large}
Technical Report -- Universidade Federal Fluminense -- May 2017
\end{large}

\end{center}

\noindent
\textbf{Abstract.} We propose an iterated local search based on several classes of local and large neighborhoods for the bin packing problem with conflicts. This problem, which combines the characteristics of both bin packing and vertex coloring, arises in various application contexts such as logistics and transportation, timetabling, and resource allocation for cloud computing. We introduce $\cO(1)$ evaluation procedures for classical local-search moves, polynomial variants of ejection chains and assignment neighborhoods, an adaptive set covering-based neighborhood, and finally a controlled use of 0-cost moves to further~diversify~the~search. The overall method produces solutions of good quality on the classical benchmark instances and scales very well with an increase of problem size. Extensive computational experiments are conducted to measure the respective contribution of each proposed neighborhood. In particular, the 0-cost moves and the large neighborhood based on set covering contribute very significantly to the search. Several research perspectives are open in relation to possible hybridizations with other state-of-the-art mathematical programming heuristics for this problem.

\vspace*{0.2cm}


\vspace*{0.5cm}

\thispagestyle{empty}
\pagenumbering{arabic}

\section{Introduction}

The Bin Packing Problem with Conflicts (BPPC) is a difficult combinatorial optimization problem which brings together the features of bin packing and vertex coloring problems.
We are given a set of bins of identical capacity $Q$, a set of items $\cI=\{1,\dots,n\}$ with weights $w_1, \dots, w_n$ and a conflict graph $G=(V,E)$. Any edge $(i,j) \in E$ represents a conflict between items $i$~and~$j$. The objective of the BPPC is to assign each item to a bin, in such a way that the number of bins in use is minimized and there are no two items in conflict in the same bin. This problem can be formulated as follows \citep{Gendreau2004}:
\begin{align}
\min \hspace*{0.2cm}  &\sum\limits_{k=1}^{n}   y_k  \label{eq:fo}\\
\text{s.t.}  \hspace{0.3cm} & \sum\limits_{i = 1}^{n} w_{i}x_{ik}  \leqslant   Q y_{k} & &  k = 1, \dots, n \label{eq:r1}  \\
& \sum\limits_{k = 1}^{n} x_{ik} = 1  & &    i  = 1, \dots, n      \label{eq:r2} \\
& x_{ik} + x_{jk} \leqslant 1              & &    (i, j) \in E, k = 1, \dots, n \label{eq:r3}\\
& y_{k} \in \{0,1\}                   & &    k = 1, \dots, n                   \label{eq:r4}\\
& x_{ik} \in \{0,1\}                  & &    i= 1, \dots, n, k = 1, \dots, n   \label{eq:r5} 
\end{align}
In this model, each binary variable $x_{ik}$ is set to $1$ if and only if item $i$ is assigned to bin $k$, and each binary variable $y_k$ is set to $1$ if and only if the bin $k$ has been used. Inequality (\ref{eq:r1}) enforces the capacity constraints, Equation (\ref{eq:r2}) states that each item should be assigned to exactly one bin, and Inequality (\ref{eq:r3}) models the conflict constraints.

The BPPC is linked to several important applications.
In the field of transportation and logistics, restrictions and compatibility constraints are imposed on several classes of products, including food, medicine or hazardous materials. Combinations of flammable, explosive, or toxic products should never be carried jointly in the same load~\citep{Hamdi-Dhaoui2014,Minh2013}, leading to difficult item-to-vehicle assignment decisions during operational planning.
The problem also arises in examination timetabling, where we search for a schedule that respects room capacities and avoids conflicts \citep{Laporte1984}. No student, for example, should take more than one exam at the same time. Finally, another application comes from the field of parallel computing, where we aim to assign a set of tasks to a minimum number of processors, subject to restrictions on the collocation of some conflicting tasks in some machines \citep{Jansen1999,Masson2012a}.

From a methodological standpoint, the problem is also of high interest.
It includes two types of constraints: capacity restrictions and conflicts. Each constraint class, alone, leads to a NP-hard subproblem: bin packing or vertex coloring. It is also noteworthy that the research on the BPPC has been in majority concentrated on approximation algorithms \citep{Epstein2008a} and mathematical-programming techniques. Mathematical programming has been successful on the BPPC up to this date, since decomposition methods such as branch-and-price \citep{Muritiba2009,Elhedhli2011,Sadykov2013a} are very efficient for problem instances with few items per bin (a common characteristic of most of the test sets available in the literature). Even in these conditions, the current mathematical programming algorithms remain applicable only to medium-scale problems, and many instances with one thousand items cannot be exactly solved to this date. This is unfortunately insufficient for several applications of BPPC, for cloud computing and large-scale task allocation problems \citep{Masson2012a}, which can involve tens or hundreds of thousands of items.

The objective of this article is to progress on the understanding and development of metaheuristics for this class of problems.
A significant part of recent research on metaheuristics, in the last five years, has been focused on producing \emph{novel} search concepts, often from natural analogies.
However, as in \cite{Sorensen2015}, we believe that part of this research is misled. Beyond natural phenomenons, the success of many metaheuristics come from a careful definition of their neighborhoods, efficient techniques to \emph{search} them, and an exploration strategy which finds a sensible balance between intensification and diversification \citep{Blum2003,Vidal2012a}. Metaheuristics are also better understood in terms of their basic operations: neighborhoods, local search, restarts and perturbations, recombinations, short and long-term memories and diversification procedures, among others. Following this general logic, we develop a simple metaheuristic framework, an iterated local search, but enhance this search procedure with a variety of local and large (exponential-sized) neighborhoods, with the aim of carefully investigating their relative contribution to the search. Overall, this leads to a simple and efficient algorithm, plus a more complete version with additional classes of neighborhoods, which both return results of good quality on classic instances for the considered problem. We also observe that set covering-based neighborhoods had the biggest impact on the search performance, as well as the proposed diversification strategies based on \emph{0-cost} moves.

Overall, the main contributions of this paper are:
\begin{enumerate}[nosep]
	\item A simple and efficient ILS approach for the BPPC;
	\item New ejection chain, assignment, and grenade neighborhoods tailored for the problem;
	\item Efficient amortized $\cO(1)$ move evaluations for standard enumerative neighborhoods;
	\item A controlled use of \emph{0-cost} moves to better explore the search place;
	\item An investigation of the relative contribution of all proposed neighborhoods. \vspace*{-0.1cm}
\end{enumerate}

\section{Related literature}
\label{sec:literature}

Although conflict constraints have a long history in the domain of timetabling \citep{Laporte1984}, the first studies dedicated to the bin packing problem with conflicts appeared in \cite{Jansen1997} and \cite{Jansen1999}. Early works were focused on approximation algorithms, a research line which was later pursued in \cite{Epstein2008} and \cite{Epstein2008a}. Unfortunately, no bounded polynomial approximation scheme can be obtained for the BPPC with general conflict graphs \citep{Jansen1999}. As such, most approximation results concern specific graph structures, such as perfect graphs, interval graphs, and bipartite graphs, with approximation ratios of $2.5$, $2.33333$, and $1.75$, respectively \citep{Epstein2008a}.

More recently, \cite{Gendreau2004} investigated some heuristics (without performance guarantee) for the problem. The authors proposed six different constructive heuristics, as well as two lower bounds, and a first set of benchmark instances based on the data set of \cite{Falkenauer1996} for the bin packing problem. Unfortunately, these instances are not available anymore to this date, and thus were re-generated later on \mbox{\citep{Muritiba2009,Elhedhli2011}}.

Subsequently, a significant research effort has been dedicated to mathematical programming algorithms for the problem. \cite{Khanafer2010} proposed new lower bounds based on the concept of data-dependent dual feasible solutions. Furthermore, three recent~articles proposed exact branch-and-price procedures \citep{Muritiba2009,Elhedhli2011,Sadykov2013a}. These articles use different branching rules, initial columns, and sub-procedures for the pricing problem (a knapsack problem with conflicts). When the conflicts form an interval graph, the pricing problem can be solved efficiently via dynamic programming \citep{Sadykov2013a}, while branch-and-bound procedures can be used in more general cases. These exact methods can solve many benchmark instances, with up to 1000 items in some cases. Still, instances with a large number of items per bin (e.g., set (da) of \citealt{Sadykov2013a}) remain challenging. The computational effort can also vary significantly from one instance to another, and becomes impracticable for large data sets with general graphs.

Metaheuristics for the BPPC have received, to this date, less attention than mathematical-programming algorithms. An advanced population-based metaheuristic was proposed in \cite{Muritiba2009}, and used to generate good upper bounds and initial columns. The method is a complex combination of a tabu search based on the impasse class neighborhoods of \cite{Morgenstern1996}, with a genetic algorithm using the crossover operator of  \cite{Galinier1999}.
\cite{Sadykov2013a} also introduce a diving heuristic based on a controlled partial exploration of the branch-and-price search tree. This method is a good alternative between exact solution approaches and heuristics, as it combines the benefits of mathematical programming for problem instances with few items per bin, with a shorter CPU time. Overall, metaheuristics for the BPPC still deserve further investigation.

Finally, as the BPPC is a generalization of both the bin packing problem and vertex coloring, results about metaheuristic searches for these two subproblems can be a good source of inspiration.
The literature on these two subproblems is extensive, and we refer to \cite{Malaguti2010,Lewis2012,Quiroz-Castellanos2015,Delorme2016} and \cite{Lewis2016} for some detailed reviews. Some of the current best metaheuristics for the bin packing problem include the grouping genetic algorithm of \cite{Falkenauer1996}, the extension of the Perturbation-MBS method by \cite{Fleszar2011}, and the grouping genetic algorithm with controlled gene transmission of \cite{Quiroz-Castellanos2015}. Population-based heuristics tend to be very frequently used, due to their capability of exploring very diverse areas of the search space. For the vertex coloring problem, some of the state-of-the-art metaheuristics include the TabuCol approach of \cite{Hertz1987}, the PartialCol algorithm of \cite{Blochliger2008} with fluctuation of the objective function, the hybrid evolutionary algorithms of \cite{Galinier1999,Malaguti2008}, and the ant colony optimization metaheuristic of \cite{Dowsland2008}.

\section{Methodology}
\label{sec:algorithm}

We opted for a simple metaheuristic framework, which allows to carefully investigate the impact of several local and large neighborhoods. Our method is thus built on an Iterated Local Search (ILS) metaheuristic, with several neighborhood classes for solution improvement, and one single perturbation procedure. Its pseudo-code is presented in Algorithm \ref{alg:ILSBPPC}.

\begin{algorithm}[htbp]
	\normalsize
	\begin{algorithmic}[1]
		\STATE  \strut $ \cS_\textsc{best} \leftarrow$ BuildInitialSolution();
		\STATE $K_\textsc{lb} \gets$ ComputeLowerBound(); 
		\WHILE{NbBins($\cS_\textsc{best}$) $\geq K_\textsc{lb}$ \textbf{and} $\cS_\textsc{best}$ is feasible}
		\STATE $\cS_\textsc{best} \gets$ ReduceNbBins($\cS_\textsc{best}$); 
		\STATE $\cS \gets \cS_\textsc{best}$;
		\STATE $i_\textsc{shak} \gets 0$;
		\WHILE{$i_\textsc{shak} \leq N_\textsc{shak}$ \textbf{and} $\cS_\textsc{best}$ is not feasible}
		\FOR{$N_\textsc{ls}$ iterations}
		\STATE $\cS \gets$ LocalSearch($\cS$);
		\STATE $\cS \gets$ AssignmentNeighborhood($\cS$);
		\STATE $\cS  \gets$ EjectionChains($\cS$);
		\STATE $\cS \gets$ Grenade($\cS$);
		\ENDFOR
		\STATE \textbf{if} $ \exists \ k \in \mathbb{N}^+ \text{ such that }  i_\textsc{shak} = k \times N_\textsc{sc}$  \textbf{then} $\cS  \gets$ SetCovering();
		\STATE \textbf{if} $cost(\cS) < cost(\cS_\textsc{best})$, \textbf{then} $\cS_\textsc{best} \gets \cS$ \textbf{and} $i_\textsc{shak} \gets 0$; \textbf{else}  $i_\textsc{shak}  \gets i_\textsc{shak} + 1$; 
		\STATE $\cS  \gets$ Shaking($\cS_\textsc{best}$);    
		\ENDWHILE
		\ENDWHILE
	\end{algorithmic}
	\caption{ \strut Iterated local search for the BPPC \label{alg:ILSBPPC}}
\end{algorithm}

An initial solution is first generated via a constructive heuristic (Line $1$). Subsequently, to decrease the number of used bins, the method iteratively removes one bin (Line $4$), randomly assigning its items to other bins, and attempts to solve the resulting conflicts or capacity excesses by means of iterations of local search with 0--cost moves, ejection chains, grenade moves, resolutions of set covering or assignment formulations, and perturbation procedures (Lines $7$--$17$). Using the terminology of the vertex coloring problem, and as surveyed in \cite{Lewis2012}, this type of strategy exploits the search space of \emph{complete, but improper} colorings. The algorithm terminates when no feasible solution is found after $N_\textsc{shak}$ perturbations, or if a trivial lower bound $K_\textsc{lb}$ on the number of bins (the sum of items weight divided by bin capacity) is reached. We now describe each component of the algorithm in further details.

\subsection{Initial Solution}
The initial solution is generated by means of the modified first fit decreasing heuristic \citep{Gendreau2004}. Items are enumerated in non-decreasing order of weight, and inserted in the first bin which has enough residual space and does not contain conflicting items. This leads to a feasible initial solution with $K$ bins.

\subsection{Local Search} 

The local search procedure aims to improve any infeasible solution produced by the removal of one bin or by the shaking operator. To define an improvement, the cost of each solution $\cS$ is evaluated as the sum of the cost of its bins $B \in \mathcal{S}$ given by Equation (\ref{solution-cost}). In this equation, $C(B)$ represents the current number of conflicts in the bin, $W(B)$ represents the total excess weight, and $\omega^\textsc{c}$ and $\omega^\textsc{w}$ are the penalty factors associated to each unit conflict and weight violation.
\begin{equation}
\Phi(B) = \omega^\textsc{c} C(B) + \omega^\textsc{w} W(B) \label{solution-cost}
\end{equation}

\noindent
\textbf{Neighborhoods and exploration.} Three classic neighborhoods are used: \textsc{Swap}, \textsc{Relocate} and \textsc{Swap2vs1}.
A \textsc{Swap} move exchanges two items from different bins. 
As suggested by its name, a \textsc{Relocate} move reassigns an item to a different bin. 
Finally, a \textsc{Swap2vs1} move exchanges a pair of consecutive items from one bin with a single item from another bin. The neighborhoods are explored as detailed in Algorithm \ref{LS-structure}, enumerating the pairs of bins in random order to test the associated moves (Lines 4--5). For each bin pair, the best improving move is applied if such a move exists (Line 9), and the local search terminates when no improving move can be found.

This evaluation policy allows to save computational effort, since it is unnecessary to re-evaluate the moves between a pair of bins when the evaluation has been done in the past without success, and if the bins have not undergone any change. Moreover, at least one bin $B$ involved in a move must satisfy $\Phi(B) > 0$ to hope for an improvement. This condition is included as an additional filter (Line 4), allowing to save a significant amount of CPU time.
Finally, during the first loop of the local search, the method accepts \emph{0-cost} moves (Line~8).  These moves do not improve the objective but contribute to diversify the search.\\

\begin{algorithm}[htbp]
	\normalsize
	\begin{algorithmic}[1]
		\STATE  \strut It$_\textsc{Loop} \gets 0$
		\WHILE{a local minimum has not been attained}
		\STATE  It$_\textsc{Loop}  \gets$   It$_\textsc{Loop}  +$ 1
		\FOR{each bin $B$ in random order such that $\Phi(B) > 0$}
		\FOR{each bin $B\,' \neq B$ in random order}
		\IF{$B$ or  $B\,'$ has been modified since the last evaluation \textbf{or}  It$_\textsc{Loop}  = 1$}
		\STATE $\Delta \gets \text{CostBestMove}(B,B\,')$
		\IF{$\{\Delta < 0\} \textbf{ or } \{\Delta \leq 0 \textbf{ and }  \text{It}_\textsc{Loop}  = 1$\}}
		\STATE $\text{ApplyBestMove}(B,B\,')$
		\ENDIF
		\ENDIF
		\ENDFOR
		\ENDFOR
		\ENDWHILE
	\end{algorithmic}
	\caption{ \strut Local search with 0-cost moves \label{LS-structure}}
\end{algorithm}

\noindent
\textbf{Efficient Move Evaluations.} The computational efficiency of the move evaluations is of critical importance, as it is the main bottleneck of most recent local search-based heuristics. Any significant reduction in the CPU time needed to evaluate moves can be translated into a direct gain in terms of number of moves which can be evaluated during one run of the algorithm, and thus into enhanced solution quality. For the BPPC, move evaluations are not trivial since they require to compute the number of conflicts in the new solution. A straightforward implementation of \textsc{Swap}, for example, between bins $B$ and $B\,'$ would require $\cO(|B| + |B\,'|)$ elementary operations per move evaluation, where $|B|$ is the current number of items in a bin $B$. This effort is due to the evaluation of conflicts between the items exchanged and those existing in the bins. This effort grows linearly with the average number of items per bin, and can be responsible for the majority of the CPU time for problem instances with a large number of items per bin.

To address this issue, we propose a pre-processing and move evaluation mechanism which supports amortized $\cO(1)$ move evaluations. We rely on an array which gives $\cO(1)$ access to the current number of conflicts $\textsc{Conf}[i][B]$ between an item $i$ and all items in bin $B$. Let $d_i$ be the degree of $i$ in the conflict graph. This array can be built in $\cO( Kn + \sum_{i=1}^n d_i )$ elementary operations for the initial solution. Subsequently, any solution modification can be assimilated to a sequence of item relocations. For each relocation of an item $i$ from a bin $B$ to a bin $B\,'$, the $\textsc{Conf}$ array is updated in $\cO(d_i)$ by means of Algorithm \ref{update-structures}.

\begin{algorithm}[htb]
	\normalsize
	\begin{algorithmic}[1]
		\FOR{\strut  each item $j$ in conflict with $i$}
		\STATE $\textsc{Conf}[j][B] =  \textsc{Conf}[j][B] - 1$
		\STATE $\textsc{Conf}[j][B\,'] =  \textsc{Conf}[j][B\,'] + 1$
		\ENDFOR
	\end{algorithmic}
	\caption{\strut  Update mechanism -- Relocation of one item $i$ from bin $B$ to $B\,'$}
	\label{update-structures}
\end{algorithm}

This data structure can now be used to perform efficient move evaluations.
Consider a \textsc{Swap} move between an item $i$ from bin $B$, with item $j$ from bin $B\,'$. The difference $\Delta$ in the number of conflicts after and before the move can be computed as follows:
\begin{equation}
\Delta =  \textsc{Conf}[i][B\,'] +  \textsc{Conf}[j][B]  -  \textsc{Conf}[i][B] - \textsc{Conf}[j][B\,'] - 2 \times  \textsc{isConflict}(i,j) \label{conflictcomputation}
\end{equation}

Remark the corrective term $2 \times  \textsc{isConflict}(i,j)$, which takes value $2$ if and only if there is a conflict between the items $i$ and $j$. This term comes from the fact that the array $\textsc{Conf}[i][B\,']$ contains the number of conflicts when moving $i$ in $B\,'$, but \emph{before} the removal of item $j$, and vice-versa for the array $\textsc{Conf}[j][B]$.

\begin{figure}[htbp]
	\centering
	\includegraphics[width=0.62 \textwidth]{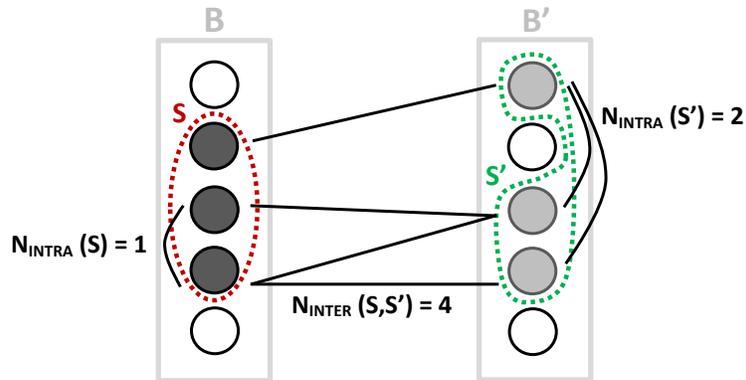}
	\caption{A generalized move involving the exchange of three items in $B$ with three items in $B\,'$.}\label{fig:move-eval}
\end{figure}

This approach can be generalized to evaluate in $\cO(1)$ any move involving an exchange of bounded sets of items $S$ and $S\,'$ between a pair of bins $B$ and $B\,'$, hence allowing efficient evaluations for all the considered moves. An example of a move, which exchanges three items of each bin, is displayed in Figure \ref{fig:move-eval}. Let $N_\textsc{inter}(S,S\,')$ be the number of conflicts between pairs of items in $S$ and $S\,'$, let $N_\textsc{intra}(S)$ be the number of conflicts within the set $S$, and let $N_\textsc{intra}(S\,')$ be the number of conflicts within the set $S\,'$. The difference in the number of conflicts after the application of the move can be obtained from Equation (\ref{conflictcomputation2}).

\begin{equation}
\begin{aligned}
\Delta &= \sum_{i \in S} ( \textsc{Conf}[i][B\,'] - \textsc{Conf}[i][B] )  + \sum_{j \in S\,'} (  \textsc{Conf}[j][B]  - \textsc{Conf}[j][B\,'] ) \\
&+ 2 \times ( N_\textsc{intra}(S) + N_\textsc{intra}(S\,') - N_\textsc{inter}(S,S\,') ) \label{conflictcomputation2}
\end{aligned}
\end{equation}

Capacity excesses can also be simply evaluated in $\cO(1)$.
Overall, this leads to move evaluations in $\cO(1)$ at the price of a slightly higher computational effort when updating the solution. In all computational experiments, we observed that hundreds of moves are tested, in average, before even applying one of them. The complexity needed to initialize and update the data structures remained largely dominated by the effort dedicated to move evaluations, leading to the announced amortized $\cO(1)$ evaluations.

\subsection{Large Neighborhoods}

Our hybrid ILS is built on simple perturbation and local search concepts, but it relies on four large neighborhoods to enhance the search performance.
The first two neighborhoods -- \textsc{Assignment} and \textsc{Ejection Chains} -- exploit restrictions of the problem which have the benefit to be combinatorial but still polynomial. This leads to neighborhoods of exponential size which can be searched in polynomial time.
The third neighborhood --  \textsc{Grenade} -- is based on the enumeration of a larger move.
Finally, the fourth neighborhood --  \textsc{Set Covering} -- uses the fact that effective integer programming solvers can address the set covering formulation of the problem in the presence of a limited pool of columns (item combinations in a bin). These solvers are used to combine high-quality columns from local minima to produce better solutions.

Some of these families of neighborhoods are well known in the operations research literature, and we refer to \cite{Deineko2000} and \cite{Ahuja2002a} for detailed surveys and analyses. Not all combinatorial optimization problems allow for an efficient search of such exponential-sized neighborhoods. For the case of the BPPC, such a search is possible. We now describe these large neighborhoods, tailored for the problem, as well as a specific form of ejection chains, which also allows a polynomial exploration of an exponential subset of solutions, under the condition that ejections between bins should follow a pre-specified ordering.

\subsubsection{Assignment Neighborhood}
\label{section-assign}
 
The assignment neighborhood selects some items for removal, and reinserts these items in the \emph{holes} thus formed by solving an assignment problem in a weighted bipartite graph. Such a methodology was successfully used for traveling salesman and vehicle routing problems in \cite{Sarvanov1981}, \cite{Gutin1999} and \cite{TothTramontani2008}.

\begin{figure}[htbp]
	\centering
	\vspace*{0.3cm}
	\includegraphics[width=0.68\textwidth]{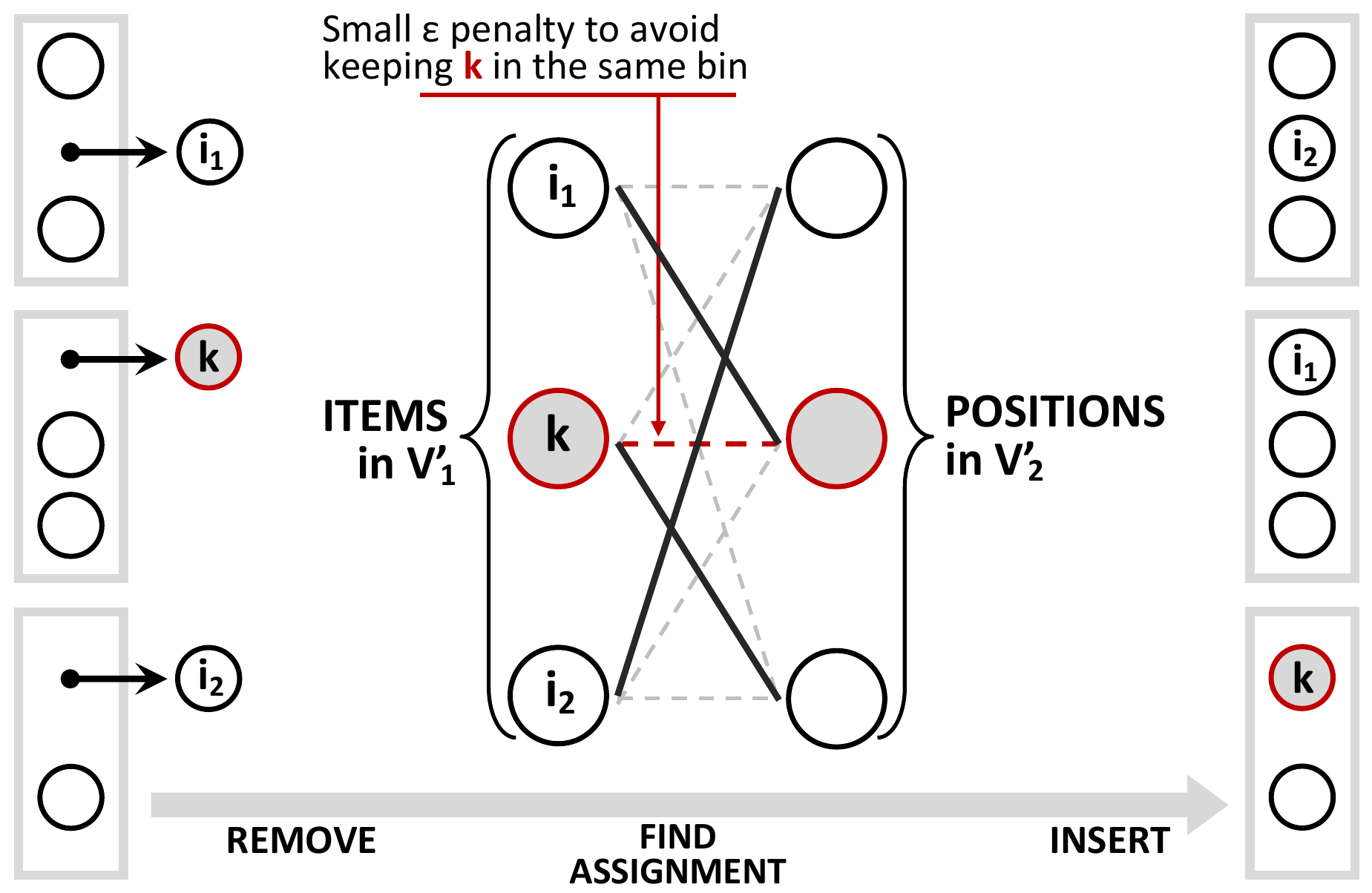}
	\caption{Assignment neighborhood and its associated bipartite graph $G'$.}\label{fig:assign}
\end{figure}

We adapt this approach to the BPPC as follows. In our context, only a few \emph{problematic} items usually present conflicts or belong to a bin with excess weight during the course of the search. To improve the solution, better allocation decisions should be sought for some of these items with simultaneous movements of other items.
Hence, the proposed algorithm randomly selects one problematic item, call it $k$, as well as $N_\textsc{assign}$ random items, at most one in each bin. Together, these items form a vertex set $V'_1$.
Let $V'_2$ be a copy of this set of vertices, and define the edge set $E' = \{ (i,j) \text{ such that } \allowbreak i \in V'_1 \text{ and } j \in V'_2 \}$. The cost of each edge $(i,j)$ is defined as the cost of inserting item $i$ in the bin $B(j)$ of item $j$ (which has been removed beforehand). Finally, to break possible ties and favor a movement of $k$, an $\epsilon$-penalty is added to the cost of the edge linking $k$ with its copy.  This defines a bipartite graph  $G'=(V'_1,V'_2,E')$, illustrated in Figure \ref{fig:assign}.

Then, finding a best re-allocation of items to bins is done by solving a linear-sum assignment problem on this graph. The optimal solution is found in $\cO(n^3)$ by means of an efficient implementation of the Hungarian algorithm~\citep{kuhn1955}. Any non-trivial optimal solution of this assignment model (different from the pairing of each node $i$ with its clone) corresponds to a move of the BPPC solution. This move is applied, leading to a new (better or equal) incumbent solution in the~ILS.

\subsubsection{Ejection Chains}

Ejection chains allow to explore a different subset of neighbor solutions obtained via chained relocations of items (see, e.g., \citealt{Thompson1993}, \citealt{Glover1996a} and \citealt{Deineko2000}).
A key difference with the previous assignment neighborhood is that the choice of items to be relocated is not a-priori fixed. On the other hand, to allow for a polynomial exploration procedure, we restrain item relocations so as to comply with a pre-defined bin ordering.

\begin{figure}[htbp]
	\centering
	\includegraphics[width=0.7\textwidth]{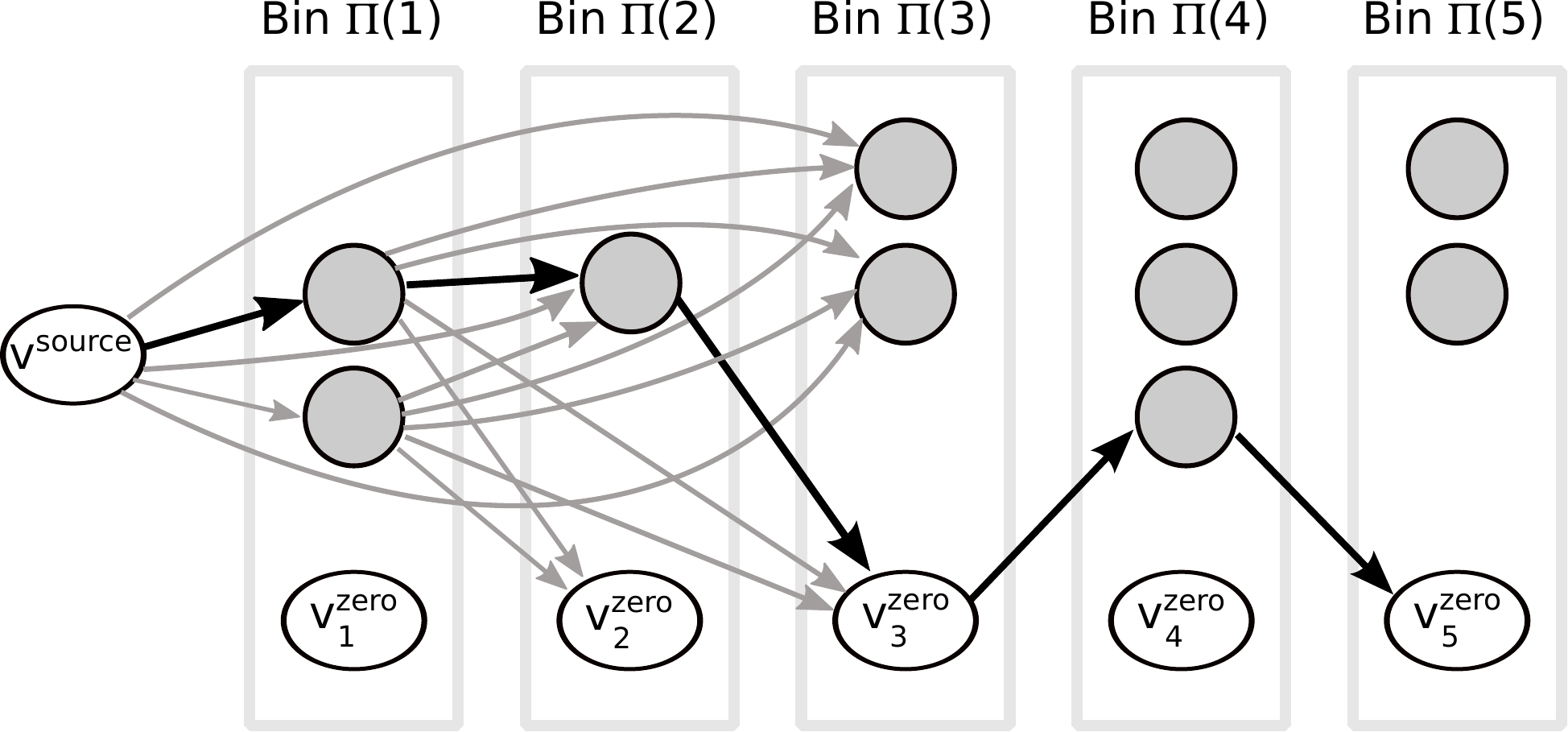}
	\caption{Ejection-chains graph, and a possible solution.}\label{fig:EChains}
\end{figure}

Our ejection chains algorithm works as follows.
First, choose a random order $\Pi$ of the bins in the current solution.
Then, define an auxiliary graph $G''=(V'',A'')$ with $V'' = \cV \cup \cV^\textsc{zero}  \cup v^\textsc{source}$, illustrated~in~Figure~\ref{fig:EChains}.
The set $\cV$ contains one node for each item, while $\cV^\textsc{zero}$ contains one node $v^\textsc{zero}_{k}$ for each bin~$k$ in the current solution.
Each node in $\cV$ will model the possible replacement of one item by another, and each node in $\cV^\textsc{zero}$ will model the possible insertion of an item and the end of an ejection chain.
The node $v^\textsc{source}$ represents a source. We denote by~$B(i)$ the bin that contains vertex $i$ in the current solution. The set $A''$ contains one arc $(i,j)$ for any pair of nodes $i \in \cV \cup \cV^\textsc{zero}$ and $j \in \cV \cup \cV^\textsc{zero}$ such that bin $B(i)$ precedes bin $B(j)$ in the order $\Pi$, and one arc $(v^\textsc{source},j)$ for any $j \in \cV$. The costs of the arcs are defined as follows:
\begin{equation}
\begin{small}
c_{ij} = 
\begin{cases} 
\text{the cost difference of bin $B(j)$ when replacing item $j$ by item $i$}  &\mbox{if } i \in \cV \text{ and } j \in \cV, \\
\text{the cost difference of bin $B(j)$ when removing item $j$}  &\mbox{if } i \in  \cV^\textsc{zero}\cup v^\textsc{source} \text{ and } j \in \cV, \\
\text{the cost difference of bin $B(j)$ when inserting item $i$}  &\mbox{if } i \in \cV \text{ and } j \in  \cV^\textsc{zero}.
\end{cases} \nonumber
\label{costs}
\end{small}
\end{equation}

A shortest path in this graph from the source to any node in $\cV^\textsc{zero}$ is equivalent to a sequence of combined item insertions and removals. Note that, thanks to the use of the nodes in $\cV^\textsc{zero}$, we can possibly obtain a collection of several disjoint ejection chains. The solution is finally updated in case of improvement.

\subsubsection{Grenade Neighborhood}

The \textsc{Grenade} neighborhood has been previously used for the vertex coloring problem in \cite{Avanthay2003}, and consists in the enumeration of some larger moves.
In our context, we consider in turn each item $k$ in the set of problematic items $\mathcal{P}$ (as defined in the previous section: with either a conflict or in a bin with excess capacity).
For each possible bin $B \neq B(k)$, the method evaluates the possibility of relocating $k$ in $B$ and jointly relocating any conflicting item $i$ of $B$ to another bin by means of a best insertion criterion. The total cost of these combined relocations is evaluated and the best grenade move for $k$ is applied in case of improvement.

\begin{figure}[htb]
	\centering
	\includegraphics[width=0.6 \textwidth]{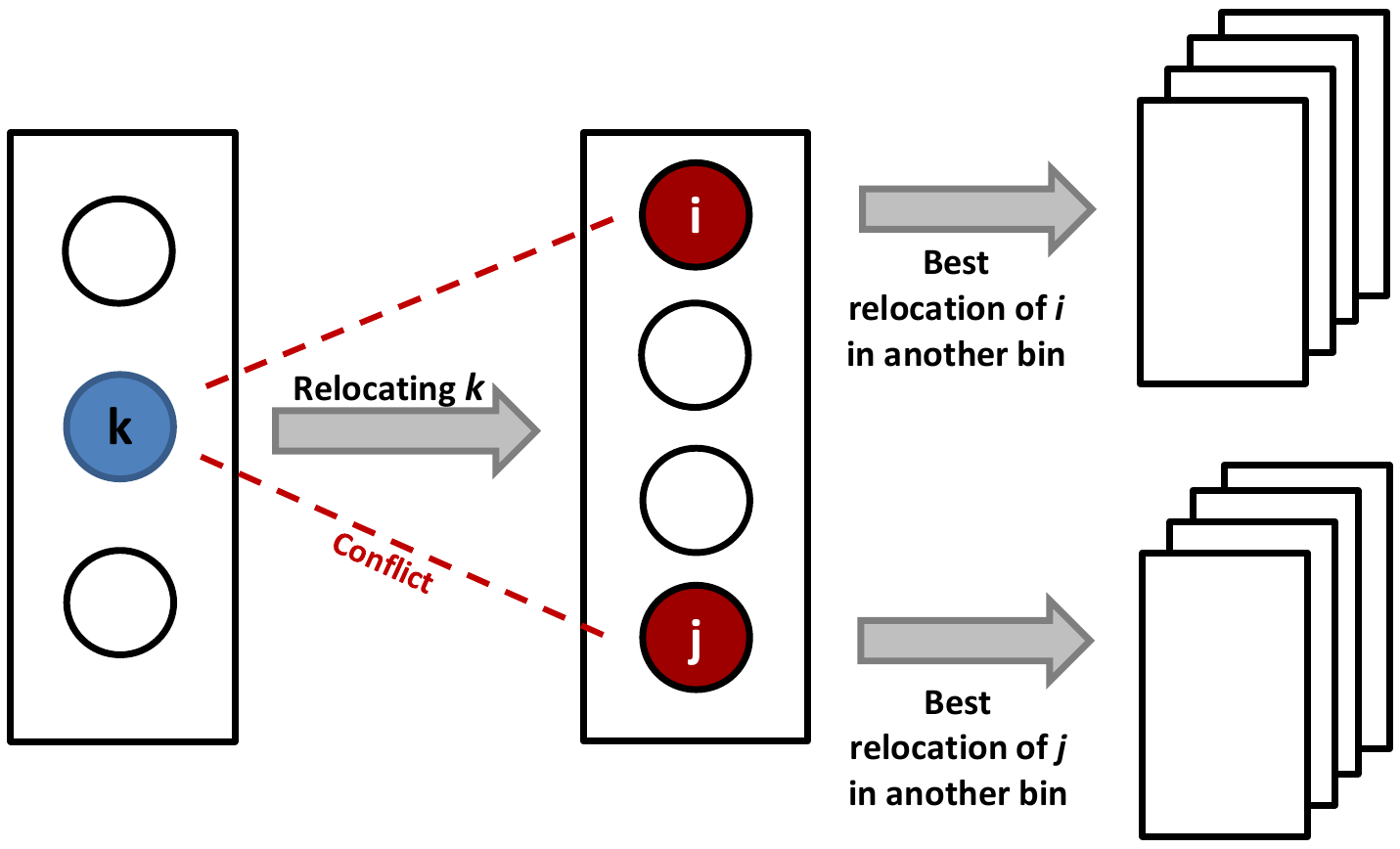}
	\caption{\textsc{Grenade} move. Relocating the item $i$ creates two conflicts which are resolved by other relocations.}\label{fig:grenade}
\end{figure}

The overall scheme is illustrated in Figure~\ref{fig:grenade}, and the pseudo-code of the neighborhood exploration is detailed in Algorithm~\ref{Grenade-structure}. Note that the search consists of a single loop over the items $k \in \mathcal{P}$, as a complete descent towards a local minimum of this neighborhood would be time consuming.

\begin{algorithm}[htbp]
	\normalsize
	\begin{algorithmic}[1]
		\FOR{\strut each item $k \in \mathcal{P}$ in random order}
		\FOR{each bin $B \neq B(k)$}
		\STATE $\Delta \gets$ Cost of a relocation of item $k$ in $B$
		\FOR{each item $i \in B$ in conflict with $k$, in random order}
		\STATE $\Delta \gets \Delta \ +$ Least-cost relocation of $i$ in a bin different from $B$
		\ENDFOR
		\STATE UpdateBestGrenadeMove($\Delta$,$k$)
		\ENDFOR
		\STATE \textbf{if} BestGrenadeMove($k$) $< 0$, \textbf{then} apply this move
		\ENDFOR
	\end{algorithmic}
	\caption{ \strut Exploration of the  \textsc{Grenade} neighborhood \label{Grenade-structure}}
\end{algorithm}

\subsubsection{Adaptive Set Covering}

Set covering (or set partitioning) formulations have been successfully used to solve a variety of combinatorial optimization problems to optimality and to generate larger neighborhoods (see, e.g.,  \citealt{Monaci2006},  \citealt{Muter2010} and \citealt{Subramanian2013a} for recent applications on a wide range of packing, coloring and routing problems). In our context, we investigate a set covering formulation in which the maximum number of bins is fixed, and one aims to minimize the sum of column costs associated to possible capacity of conflict constraints violations, and use it as a base for an additional large neighborhood in the ILS. To limit the computational effort, this procedure is used only a few times during the search, after each $N_\textsc{sc}$ consecutive iterations of the local search procedure without improvement.

The considered set covering formulation is expressed in Equations (\ref{eq:foSP}--\ref{eq:r3SP}).
It involves the set $R$ of all possible columns, where each column represents a combination of items, associated with a cost $c_j$.
The cost $c_j$ is based on the same weighted penalties for conflicts and capacity excess as in Equation (\ref{solution-cost}), and let $R_i \subseteq R$ be the subset of columns containing an item $i \in \cV$. The goal of the formulation is to find a set of up to $K$ columns that covers all items with minimum cost. We highlight the fact that the positive-cost columns (with $c_i > 0$) correspond to sets of items with excess capacity or conflicts, and that any solution with cost $0$ would correspond to a feasible packing in $K$ bins. \begin{align}
 \min \hspace*{0.2cm} &\sum\limits_{j \in R} c_{j}u_{j}                        &                         \label{eq:foSP} \\
\text{s.t.}  \hspace{0.3cm}    & \sum_{j \in R_i} u_{j}  \geqslant   1         &     \forall i \in \cV   \label{eq:r1SP} \\
& \sum_{j \in R} u_{j} \leqslant   K                                           &                         \label{eq:r2SP} \\
& u_{j} \in \{0,1\}                                                            &     \forall j \in R     \label{eq:r3SP} 
\end{align}

This formulation involves an exponential number of decision variables ($2^n - 1$). To circumvent this issue, we restrict the model to a subset of columns $R' \subset R$, obtained from local minima of the ILS.
More specifically, before every shaking iteration, the columns of the current solution are added to the pool $R'$, filtering possible duplicates (this can easily implemented by hashing), and controlling the size of the pool to remain below a limit $S_\textsc{pool}$. The new columns start to replace the oldest columns when this limit is attained.
The formulation is solved by means of an integer programming solver subject to a CPU time limit $T_\textsc{Limit}$. The pool-size limit is adapted during the search, with the aim to balance the difficulty of the problem and thus obtain good solutions with a formulation that remains tractable.
If the model was solved to the optimality during the allowed time, then $S_\textsc{pool}$ is increased by $15\%$. On the contrary, if the time limit was attained without an optimal solution, then $S_\textsc{pool}$ is reduced by $15\%$. The optimal solution provided by this model is taken as new incumbent solution.

\subsection{Perturbation}  

Shaking is only operated after $N_\textsc{ls}$ consecutive iterations of the $0$-cost moves, local and large neighborhood search without improvement of the best solution. Indeed, the \emph{0-cost} moves themselves already allow significantly modifications to the solutions, which help to progress towards different areas of the search space. 

During shaking, $N_\textsc{shak}$ items are randomly relocated, considering $50\%$ items with conflicts or capacity excess, and $50\%$ random items.
The size of the shaking operator is a key parameter for the search.
Its value has been defined as a result of our preliminary computational experiments, as described in the next section.

\section{Computational Experiments}
\label{sec:results}

The goal of these experiments is to evaluate the performance of the proposed method on the BPPC, as well as to identify the relative contribution of each neighborhood and the impact of its key design choices to drive future research. The classical benchmark instances for the BPPC are divided into six sets: (t) and (u) from \cite{Muritiba2009}, and (ta), (ua), (d), (da) from \cite{Sadykov2013a}, for a total of 2060 instances ranging from 60 to~1000~items.

The instances of the group (t -- \emph{triplet}) and (u -- \emph{uniform}) were first derived from the bin packing instances of \cite{Falkenauer1996} by \citet{Gendreau2004}. 
The generation procedure, however, contains some randomized choices and the original instances are no longer available. Thus, these instances were generated again in \cite{Muritiba2009} and \cite{Elhedhli2011}. In this paper, we rely on the files of \cite{Muritiba2009}, as this allows a comparison with the only existing population-based metaheuristic in the literature, as well as the current state-of-the-art mathematical programming -- heuristic and exact -- algorithms of \cite{Sadykov2013a}.
The set (t) includes four groups of instances with a number of items ranging from 60~to~501. One remarkable characteristic of this instance group is that, in the optimal bin packing solution, without conflict constrains, the bins are filled by exactly three items each. In a similar manner, the set (u) includes four classes of instances, with 120 to 1000 items. In both groups, the conflict graph is an interval graph, with a density ranging from 0\%~to~90\%. Previous articles did not always report results for the instances with 0\% density, leading to 360 instances per set (9 density levels $\times$ 4 size levels $\times$ 10 instances).

The instances of sets (ta) and (ua) have the same characteristics as the instances of sets (t)~and~(u), but with an arbitrary conflict graph. These instances were generated by \cite{Sadykov2013a} with edge densities from 10\% to 90\%, for a total of 360 instances in each set. Finally, the instances of sets (d) and (da) were generated with item weights uniformly distributed in \mbox{[500, 2500]}, and a bin capacity of 10000. These instances include either 120, 250 or 500 items, with interval~(d) or arbitrary~(da) conflict graph. The objective of these instances was to include a higher average number of items per bin and obtain more difficult problems for column generation-based methods. This objective was, however, only partially attained, since the average number of items per bin in the best known solutions remain relatively small for all data sets: 2.18, 1.96, 2.91, 2.48, 2.76, and 5.83 for the classes (t), (u), (ta), (ua), (d), and (da), respectively. For this reason, these instance sets remain ideal for existing branch-and-price methods.%as the pricing algorithm can quickly find improving columns, leading to a high performance which is difficult to match by means of neighborhood-centered metaheuristics.

\subsection{Performance Evaluations}
\label{Performance}

The first experiment aims to measure the performance of the proposed ILS with large neighborhoods relatively to the performance of the best current metaheuristic, the population heuristic~(PH) of \cite{Muritiba2009}, and the best mathematical-programming based heuristic, the diving approach with limited discrepancy search (DH-LDS) of \citet{Sadykov2013a}.
As the total number of instances for the BPPC is large, the results are usually presented in an aggregated form. For a precise comparison, the detailed results of PH and DH-DLS  were kindly provided by the authors in a personal communication. The proposed algorithm was coded in C++, uses CPLEX 12.5.1, and was run on a single thread of a Xeon X5675, 3.07 GHz CPU. PH and DH-LDS were run on Pentium IV 3.0GHz and Xeon X5460 3.16GHz CPUs, respectively.

A preliminary calibration of the search parameters -- size and frequency of the shaking operator, number of iterations of the method, size of the set covering subproblem  -- was conducted with the aim of producing good solutions in a CPU time which is comparable with previous research for medium-sized instances. This led to a standard set of parameters:  $N_\textsc{shak}= 50$, $N_\textsc{ls}= 100$, $N_\textsc{sc}=25$, $S_\textsc{pool} = 1500$, $T_\textsc{Limit} = 20$ and $S_\textsc{shak}=3$. The impact of any deviation from these search parameters will be investigated in deeper details in Section~\ref{Analysis}.

The proposed method was then run 10 times for each instance, with different random seeds, in order to measure its average performance. The complete version of the algorithm will be called HILS-\emph{Complete}. To examine the impact of the large neighborhood search on solution quality, we will also report the results of a simplified configuration, ILS-\emph{Simple}, which corresponds to the same algorithm without the large neighborhoods. These experiments are presented in Tables \ref{tab:sumaryNumberItems}--\ref{tab:sumaryByDensity}. In Table \ref{tab:sumaryNumberItems}, the results are aggregated for each instance class $\times$ number of items, while Table \ref{tab:sumaryByDensity} aggregates the results per instance class $\times$ density. The following metrics are reported:
\begin{itemize}[nosep,leftmargin=1.3cm]
	\item [\textbf{Gap} --] The average gap from the best known solutions (BKS) in the literature. For each instance, the gap is calculated as $100(z - z_{BKS})/z_{BKS}$, where $z$ is the value of the solution found by the method and $z_{BKS}$ is the value of the BKS;
	\item [\textbf{T(s)} --] The average CPU time per instance until termination of the search;
	\item [\textbf{Opt} --] The average number of optimal solutions found for this group of instances in one run.
\end{itemize}
For each group of instances, the best results are highlighted in boldface. The last two columns indicate the average number of bins in the best known solutions of the literature, and the number of known optimal solutions for the group. \\

\begin{table}[htb]
\centering
	\renewcommand{\arraystretch}{1.1}
	\setlength{\tabcolsep}{0.08cm}
	\scalebox{0.8}
	{
		\begin{tabular}{|M{0.8cm}M{0.8cm}|M{1cm}M{1cm}M{1cm}H|M{1cm}M{1cm}M{1cm}H|M{1cm}M{1cm}M{1cm}|M{1cm}M{1cm}M{1cm}|M{1.2cm}c|} 
			\hline
			\multicolumn{2}{|c|}{Instance} & \multicolumn{4}{c|}{HILS-\emph{Complete}} & \multicolumn{4}{c|}{ILS-\emph{Simple}} & \multicolumn{3}{c|}{PH} & \multicolumn{3}{c|}{DH-LDS} & \multicolumn{2}{c|}{BKS}\\
			Class & $n$ &  Gap  & T(s) & Opt &  Best Gap  & Gap & T(s) & Opt & Gap & Gap & T(s) & Opt &  Gap & T(s) & Opt & Bins & Opt \\ \hline
			\multirow{4}{*}{t}  & 60 &  \textbf{0.00} & 1.26 & 90.0 &  \textbf{0.00} & 0.42 & 0.72 & 82.1 & 0.22 & 0.50 & 41.68 & 81 &  \textbf{0.00} & 0.20 & 90 & 33.40 & 90 \\
			& 120 &  \textbf{0.01} & 5.17 & 89.6 &  \textbf{0.00} & 0.63 & 2.34 & 66.3 & 0.39 & 0.66 & 44.44 & 66 &  0.03 & 1.36 & 89 & 66.11 & 90 \\
			& 249 &  0.08  & 17.27 & 84.3 &  0.05  & 0.35 & 5.09 & 57.5 & 0.15 & 0.44 & 57.61 & 58 &  \textbf{0.00} & 7.81 & 90& 135.83 & 90 \\
			& 501 &  0.01  & 23.88 & 88.9 &  \textbf{0.00} & 0.26 & 13.19 & 47.9 & 0.10 & 0.23 & 65.48 & 60 &  \textbf{0.00} & 50.43 & 90 & 275.69 & 90 \\ \hline
			\multirow{4}{*}{u} & 120 &  0.03  & 3.67 & 88.8 &  \textbf{0.00} & 0.06 & 1.22 & 87.1 & 0.02 & 0.11 & 24.58 & 85 &  \textbf{0.00} & 0.76 & 90 & 70.38 & 90 \\
			& 250 &  0.01  & 17.57 & 88.8 &  \textbf{0.00} & 0.13 & 4.67 & 74.9 & 0.03 & 0.19 & 49.81 & 71 &  \textbf{0.00} & 3.34 & 90 & 143.72 & 90 \\
			& 500 &  0.01  & 37.56 & 86.8 &  0.01  & 0.37 & 16.47 & 45.8 & 0.20 & 0.18 & 66.44 & 65 &  \textbf{0.00} & 18.74 & 90 & 286.03 & 90 \\
			& 1000 &  0.01  & 90.07 & 85.7 &  0.01  & 0.53 & 65.07 & 33.6 & 0.34 & 0.20 & 105.34 & 56 &  \textbf{0.00} & 116.71 & 90 & 571.88 & 90 \\ \hline
			\multirow{4}{*}{ta} & 60 &  0.00  & 1.85 & 90.0 &  \textbf{0.00} & 0.09 & 0.78 & 88.4 & 0.06 & -- & -- & -- &  \textbf{0.00} & 0.16 & 90 & 21.87 & 90 \\
			& 120 &  \textbf{0.05}  & 23.34 & 88.1 &  \textbf{0.00} & 1.18 & 3.41 & 49.4 & 1.01 & -- & -- & -- &  \textbf{0.05} & 1.46 & 88 & 41.39 & 90 \\
			& 249 &  0.49  & 75.17 & 51.3 &  0.32  & 0.94 & 12.88 & 42.3 & 0.77 & -- & -- & -- &  \textbf{0.29} & 23.67 & 67 & 83.80 & 88 \\
			& 501 &  0.40  & 128.30 & 44.5 &  0.35  & 0.57 & 32.87 & 44.1 & 0.53 & -- & -- & -- &  \textbf{0.15} & 271.33 & 48 & 167.61 & 70 \\ \hline
			\multirow{4}{*}{ua} & 120 &  \textbf{0.00} & 3.26 & 90.0 &  \textbf{0.00} & 0.52 & 1.37 & 66.8 & 0.38 & -- & -- & -- &  \textbf{0.00} & 0.73 & 90 & 49.27 & 90 \\
			& 250 &  \textbf{0.02}  & 28.70 & 87.1 &  \textbf{0.00} & 0.55 & 6.83 & 58.9 & 0.48 & -- & -- & -- &  \textbf{0.02} & 6.21 & 87 & 100.64 & 89 \\
			& 500 &  0.08  & 154.09 & 69.8 &  0.05  & 0.38 & 26.62 & 60.8 & 0.35 & -- & -- & -- &  \textbf{0.01} & 63.10 & 80 & 200.77 & 82 \\
			& 1000 &  0.15  & 399.37 & 67.8 &  0.06  & 0.27 & 55.74 & 67.6 & 0.23 & -- & -- & -- &  \textbf{0.01} & 516.75 & 79 & 400.03 & 82 \\ \hline
			\multirow{3}{*}{d} & 120 &  \textbf{0.00}  & 3.17 & 90.0 &  \textbf{0.00} & 0.04 & 1.01 & 88.1 & 0.00 & -- & -- & -- &  -- & -- & -- & 61.82 & 90 \\
			& 250 &  \textbf{0.00}  & 11.57 & 89.9 &  \textbf{0.00} & 0.08 & 2.77 & 80.6 & 0.03 & -- & -- & -- &  -- & -- & -- & 127.93 & 90 \\
			& 500 &  \textbf{0.00}  & 23.35 & 89.5 &  \textbf{0.00} & 0.10 & 8.96 & 69.2 & 0.02 & -- & -- & -- &  -- & -- & -- & 252.79 & 90 \\ \hline
			\multirow{3}{*}{da} & 120 &  0.80  & 12.91 & 57.7 &  0.18  & 0.98 & 1.18 & 55.0 & 0.39 & -- & -- & -- &  \textbf{0.12} & 6.02 & 63 & 23.63 & 67 \\
			& 250 &  1.14  & 25.50 & 40.5 &  0.52  & 1.32 & 4.88 & 38.0 & 0.73 & -- & -- & -- &  \textbf{0.04} & 59.60 & 49 & 44.66 & 50 \\
			& 500 &  1.83  & 53.31 & 37.1 &  1.38  & 1.85 & 18.12 & 37.0 & 1.41 & -- & -- & -- &  \textbf{0.10} & 541.48 & 47 & 84.12 & 49 \\  \hline 
		\end{tabular}
	}
	\caption{Comparison of th current state-of-the-art BPPC heuristics for each instance set and size $n$.}
	\label{tab:sumaryNumberItems}
\end{table}
 
\begin{table}[htbp]
	\centering
        \vspace*{-0.5cm}
	\renewcommand{\arraystretch}{1.02}
	\setlength{\tabcolsep}{0.14cm}
	\scalebox{0.8}
	{
		\begin{tabular}{|M{0.8cm}M{0.8cm}|M{1cm}M{1cm}M{1cm}H|M{1cm}M{1cm}M{1cm}H|M{1cm}M{1cm}M{1cm}|M{1cm}M{1cm}M{1cm}|M{1cm}c|} 
			\hline
			\multicolumn{2}{|c|}{Instance} & \multicolumn{4}{c|}{HILS-\emph{Complete}} & \multicolumn{4}{c|}{ILS-\emph{Simple}} & \multicolumn{3}{c|}{PH} & \multicolumn{3}{c|}{DH-LDS} & \multicolumn{2}{c|}{BKS}\\
			Class & $\rho$ &  Gap  & T(s) & Opt &  Best Gap  &  Gap  & T(s) & Opt &  Best Gap  & Gap  & T(s) & Opt &  Gap  & T(s) & Opt & Bins & Opt \\ \hline
			\multirow{10}{*}{t}  &  0  &  \textbf{0.00}  &  0.93  &  40.0  &  \textbf{0.00}  &  1.44  &  0.97  &  25.3  &  0.75  &  \textbf{0.00}  &  0.00  &  40  &  --  &  --  &  --  &  77.50  &  40  \\  
			&  10  &  \textbf{0.00}  &  1.05  &  40.0  &  \textbf{0.00}  &  1.26  &  1.12  &  26.1  &  0.69  &  2.08  &  134.03  &  2  &  \textbf{0.00}  &  11.93  &  40  &  77.55  &  40  \\  
			&  20  &  \textbf{0.00}  &  1.77  &  40.0  &  \textbf{0.00}  &  0.72  &  1.67  &  29.4  &  0.50  &  1.20  &  131.98  &  9  &  \textbf{0.00}  &  21.93  &  40  &  77.73  &  40  \\  
			&  30  &  0.21  &  19.99  &  32.8  &  0.12  &  0.45  &  7.88  &  27.2  &  0.34  &  0.71  &  115.05  &  18  &  \textbf{0.06}  &  24.15  &  39  &  78.33  &  40  \\  
			&  40  &  \textbf{0.00}  &  8.78  &  40.0  &  \textbf{0.00}  &  0.23  &  4.09  &  30.0  &  0.07  &  0.14  &  24.55  &  36  &  \textbf{0.00}  &  19.30  &  40  &  94.10  &  40  \\  
			&  50  &  \textbf{0.00}  &  11.35  &  40.0  &  \textbf{0.00}  &  0.29  &  5.00  &  27.9  &  0.12  &  \textbf{0.00}  &  12.73  &  40  &  \textbf{0.00}  &  18.69  &  40  &  117.93  &  40  \\  
			&  60  &  \textbf{0.00}  &  13.10  &  40.0  &  \textbf{0.00}  &  0.28  &  5.87  &  28.3  &  0.08  &  \textbf{0.00}  &  21.98  &  40  &  \textbf{0.00}  &  14.18  &  40  &  141.53  &  40  \\  
			&  70  &  \textbf{0.00}  &  15.85  &  40.0  &  \textbf{0.00}  &  0.27  &  6.89  &  23.6  &  0.10  &  \textbf{0.00}  &  16.10  &  40  &  \textbf{0.00}  &  10.31  &  40  &  164.33  &  40  \\  
			&  80  &  \textbf{0.00}  &  16.94  &  40.0  &  \textbf{0.00}  &  0.15  &  7.36  &  29.3  &  0.03  &  \textbf{0.00}  &  10.18  &  40  &  \textbf{0.00}  &  8.57  &  40  &  187.78  &  40  \\  
			&  90  &  \textbf{0.00}  &  18.22  &  40.0  &  \textbf{0.00}  &  0.08  &  8.17  &  32.0  &  0.02  &  \textbf{0.00}  &  4.15  &  40  &  \textbf{0.00}  &  5.51  &  40  &  210.58  &  40  \\  \hline
			\multirow{10}{*}{u}  &  0  &  \textbf{0.00}  &  0.56  &  40.0  &  \textbf{0.00}  &  \textbf{0.00}  &  0.26  &  40.0  &  \textbf{0.00}  &  0.31  &  75.95  &  18  &  --  &  --  &  --  &  188.53  &  40  \\  
			&  10  &  \textbf{0.00}  &  0.60  &  40.0  &  \textbf{0.00}  &  \textbf{0.00}  &  0.28  &  40.0  &  0.00  &  0.11  &  57.33  &  31  &  \textbf{0.00}  &  29.43  &  40  &  188.53  &  40  \\  
			&  20  &  \textbf{0.00}  &  0.60  &  40.0  &  \textbf{0.00}  &  \textbf{0.00}  &  0.27  &  40.0  &  0.00  &  0.10  &  57.78  &  32  &  \textbf{0.00}  &  31.63  &  40  &  188.53  &  40  \\  
			&  30  &  0.05  &  2.65  &  38.7  &  \textbf{0.00}  &  0.12  &  1.32  &  36.9  &  0.07  &  0.27  &  76.15  &  25  &  \textbf{0.00}  &  37.26  &  40  &  188.53  &  40  \\  
			&  40  &  0.09  &  58.14  &  34.1  &  0.02  &  0.39  &  20.12  &  24.9  &  0.21  &  0.79  &  91.95  &  14  &  \textbf{0.00}  &  57.83  &  40  &  194.03  &  40  \\  
			&  50  &  \textbf{0.00}  &  41.99  &  39.1  &  \textbf{0.00}  &  0.62  &  22.43  &  18.2  &  0.40  &  0.10  &  46.80  &  34  &  \textbf{0.00}  &  50.78  &  40  &  236.80  &  40  \\  
			&  60  &  \textbf{0.00}  &  44.72  &  39.9  &  \textbf{0.00}  &  0.36  &  27.02  &  21.6  &  0.17  &  0.06  &  61.93  &  34  &  \textbf{0.00}  &  34.85  &  40  &  284.80  &  40  \\  
			&  70  &  0.01  &  62.13  &  38.8  &  \textbf{0.00}  &  0.45  &  35.54  &  18.0  &  0.23  &  0.06  &  64.40  &  32  &  \textbf{0.00}  &  28.84  &  40  &  330.43  &  40  \\  
			&  80  &  \textbf{0.00}  &  68.64  &  39.5  &  \textbf{0.00}  &  0.35  &  42.03  &  18.0  &  0.19  &  0.03  &  59.70  &  36  &  \textbf{0.00}  &  23.78  &  40  &  376.73  &  40  \\  
			&  90  &  \textbf{0.00}  &  55.48  &  40.0  &  \textbf{0.00}  &  0.16  &  47.72  &  23.8  &  0.07  &  0.00  &  37.88  &  39  &  \textbf{0.00}  &  19.57  &  40  &  423.68  &  40  \\  \hline
			\multirow{9}{*}{ta}  &  10  &  \textbf{0.00}  &  1.52  &  40.0  &  \textbf{0.00}  &  0.74  &  1.42  &  29.2  &  0.63  &  --  &  --  &  --  &  \textbf{0.00}  &  51.24  &  40  &  77.73  &  40  \\  
			&  20  &  \textbf{0.00}  &  2.46  &  40.0  &  \textbf{0.00}  &  0.57  &  1.84  &  30.9  &  0.56  &  --  &  --  &  --  &  \textbf{0.00}  &  42.81  &  40  &  77.78  &  40  \\  
			&  30  &  0.33  &  16.99  &  30.9  &  \textbf{0.00}  &  0.76  &  6.73  &  23.4  &  0.53  &  --  &  --  &  --  &  \textbf{0.11}  &  72.27  &  36  &  77.80  &  40  \\  
			&  40  &  0.28  &  34.60  &  30.5  &  \textbf{0.27}  &  0.28  &  13.58  &  30.1  &  0.27  &  --  &  --  &  --  &  \textbf{0.27}  &  112.75  &  28  &  78.03  &  40  \\  
			&  50  &  \textbf{0.42}  &  74.85  &  21.0  &  \textbf{0.42}  &  \textbf{0.42}  &  25.38  &  21.0  &  0.42  &  --  &  --  &  --  &  \textbf{0.42}  &  120.84  &  21  &  78.03  &  40  \\  
			&  60  &  \textbf{0.06}  &  93.88  &  32.0  &  \textbf{0.06}  &  \textbf{0.06}  &  20.04  &  32.0  &  \textbf{0.06}  &  --  &  --  &  --  &  \textbf{0.06}  &  80.08  &  32  &  78.40  &  36  \\  
			&  70  &  \textbf{0.00}  &  107.48  &  32.0  &  \textbf{0.00}  &  0.10  &  16.37  &  30.9  &  \textbf{0.00}  &  --  &  --  &  --  &  \textbf{0.00}  &  86.29  &  32  &  78.50  &  32  \\  
			&  80  &  0.44  &  82.58  &  20.3  &  0.33  &  1.66  &  16.72  &  10.5  &  1.44  &  --  &  --  &  --  &  \textbf{0.12}  &  9.40  &  38  &  79.00  &  40  \\  
			&  90  &  0.59  &  100.13  &  27.2  &  0.43  &  1.66  &  10.27  &  16.2  &  1.41  &  --  &  --  &  --  &  \textbf{0.13}  &  91.72  &  26  &  82.75  &  30  \\  \hline
			\multirow{9}{*}{ua}  &  10  &  0.01  &  2.46  &  39.3  &  \textbf{0.00}  &  0.02  &  1.14  &  39.0  &  0.00  &  --  &  --  &  --  &  \textbf{0.00}  &  74.71  &  40  &  187.58  &  40  \\  
			&  20  &  0.01  &  0.81  &  39.7  &  \textbf{0.00}  &  0.01  &  0.48  &  39.6  &  0.00  &  --  &  --  &  --  &  \textbf{0.00}  &  74.44  &  40  &  188.35  &  40  \\  
			&  30  &  \textbf{0.00}  &  0.71  &  40.0  &  \textbf{0.00}  & \textbf{0.00}  &  0.40  &  40.0  &  0.00  &  --  &  --  &  --  &  \textbf{0.00}  &  69.08  &  40  &  187.90  &  40  \\  
			&  40  &  \textbf{0.00}  &  3.69  &  39.7  &  \textbf{0.00}  &  0.06  &  1.79  &  38.7  &  0.05  &  --  &  --  &  --  &  \textbf{0.00}  &  69.72  &  40  &  186.78  &  40  \\  
			&  50  &  \textbf{0.00}  &  4.07  &  40.0  &  \textbf{0.00}  &  0.05  &  1.72  &  39.0  &  0.05  &  --  &  --  &  --  &  \textbf{0.00}  &  66.68  &  40  &  186.93  &  40  \\  
			&  60  &  0.02  &  13.52  &  38.6  &  0.01  &  0.18  &  8.54  &  34.1  &  0.17  &  --  &  --  &  --  &  \textbf{0.00}  &  81.51  &  40  &  187.03  &  40  \\  
			&  70  &  0.10  &  75.84  &  30.7  &  0.06  &  0.54  &  39.83  &  18.0  &  0.49  &  --  &  --  &  --  &  \textbf{0.03}  &  106.85  &  38  &  187.48  &  39  \\  
			&  80  &  0.18  &  306.37  &  21.8  &  0.13  &  1.08  &  86.13  &  2.0  &  0.83  &  --  &  --  &  --  &  \textbf{0.00}  &  171.81  &  34  &  187.75  &  34  \\  
			&  90  &  0.26  &  909.73  &  24.9  &  \textbf{0.04}  &  1.91  &  63.73  &  3.7  &  1.64  &  --  &  --  &  --  &  \textbf{0.07}  &  605.47  &  24  &  189.33  &  30  \\  \hline
			\multirow{9}{*}{d}  &  10  &  \textbf{0.00}  &  0.00  &  30.0  &  \textbf{0.00}  &  \textbf{0.00}  &  0.00  &  30.0  &  \textbf{0.00}  &  --  &  --  &  --  &  --  &  --  &  --  &  43.90  &  30  \\  
			&  20  &  \textbf{0.00}  &  5.22  &  30.0  &  \textbf{0.00}  &  0.06  &  2.52  &  28.7  &  \textbf{0.00}  &  --  &  --  &  --  &  --  &  --  &  --  &  58.67  &  30  \\  
			&  30  &  0.01  &  9.31  &  29.5  &  \textbf{0.00}  &  0.13  &  3.13  &  26.1  &  0.04  &  --  &  --  &  --  &  --  &  --  &  --  &  87.67  &  30  \\  
			&  40  &  \textbf{0.00}  &  10.70  &  30.0  &  \textbf{0.00}  &  0.17  &  3.57  &  25.8  &  0.03  &  --  &  --  &  --  &  --  &  --  &  --  &  114.30  &  30  \\  
			&  50  &  \textbf{0.00}  &  13.59  &  30.0  &  \textbf{0.00}  &  0.02  &  4.05  &  28.6  &  \textbf{0.00}  &  --  &  --  &  --  &  --  &  --  &  --  &  147.63  &  30  \\  
			&  60  &  \textbf{0.00}  &  16.14  &  30.0  &  \textbf{0.00}  &  0.11  &  4.82  &  24.0  &  0.05  &  --  &  --  &  --  &  --  &  --  &  --  &  174.07  &  30  \\  
			&  70  &  \textbf{0.00}  &  17.90  &  29.9  &  \textbf{0.00}  &  0.11  &  5.81  &  22.6  &  0.01  &  --  &  --  &  --  &  --  &  --  &  --  &  204.40  &  30  \\  
			&  80  &  \textbf{0.00}  &  19.58  &  30.0  &  \textbf{0.00}  &  0.03  &  6.50  &  26.8  &  \textbf{0.00}  &  --  &  --  &  --  &  --  &  --  &  --  &  235.07  &  30  \\  
			&  90  &  \textbf{0.00}  &  21.82  &  30.0  &  \textbf{0.00}  &  0.04  &  7.83  &  25.3  &  \textbf{0.00}  &  --  &  --  &  --  &  --  &  --  &  --  &  261.93  &  30  \\  \hline
			\multirow{9}{*}{da}  &  10  &  \textbf{0.00}  &  0.00  &  30.0  &  \textbf{0.00}  &  \textbf{0.00}  &  0.00  &  30.0  &  \textbf{0.00}  &  --  &  --  &  --  &  0.04  &  92.49  &  29  &  43.90  &  30  \\  
			&  20  &  \textbf{0.00}  &  0.00  &  30.0  &  \textbf{0.00}  &  \textbf{0.00}  &  0.00  &  30.0  &  \textbf{0.00}  &  --  &  --  &  --  &  0.09  &  84.50  &  29  &  43.80  &  30  \\  
			&  30  &  0.38  &  3.29  &  28.0  &  0.38  &  0.38  &  0.22  &  28.0  &  0.38  &  --  &  --  &  --  &  \textbf{0.00}  &  36.71  &  30  &  44.17  &  30  \\  
			&  40  &  0.55  &  18.54  &  23.6  &  0.27  &  0.69  &  4.09  &  22.4  &  0.55  &  --  &  --  &  --  &  \textbf{0.00}  &  79.81  &  28  &  44.03  &  28  \\  
			&  50  &  1.97  &  67.05  &  4.6  &  1.06  &  2.40  &  10.05  &  1.6  &  1.46  &  --  &  --  &  --  &  \textbf{0.00}  &  81.00  &  24  &  44.43  &  24  \\  
			&  60  &  2.30  &  74.63  &  0.0  &  1.38  &  2.65  &  12.14  &  0.0  &  1.59  &  --  &  --  &  --  &  \textbf{0.04}  &  384.41  &  2  &  45.50  &  3  \\  
			&  70  &  3.12  &  47.54  &  2.1  &  1.77  &  3.27  &  15.25  &  2.0  &  1.94  &  --  &  --  &  --  &  \textbf{0.04}  &  543.40  &  3  &  49.77  &  3  \\  
			&  80  &  2.43  &  35.15  &  7.0  &  1.36  &  2.52  &  16.01  &  6.0  &  1.62  &  --  &  --  &  --  &  \textbf{0.21}  &  348.24  &  7  &  60.30  &  8  \\  
			&  90  &  0.55  &  28.96  &  10.0  &  \textbf{0.01}  &  0.57  &  14.78  &  10.0  &  0.05  &  --  &  --  &  --  &  \textbf{0.34}  &  170.71  &  7  &  81.33  &  10  \\  \hline
			
		\end{tabular}
	}
	\caption{Comparison of th current state-of-the-art BPPC heuristics for each instance set and density $\rho$.}
	\label{tab:sumaryByDensity}
\end{table}

Two factors are determining for the performance of an optimization approach: the quality of its solutions, and the scalability of the method, i.e., how fast the CPU time grows for larger problem instances.
From these results, the following observations can be made.

In terms of solution quality, HILS-\emph{Complete} provides solutions of higher quality than the current best metaheuristic (PH), and slightly lower quality than the column generation-based diving algorithm (DH-LDS). The magnitude of the differences depends on the sets of instances. For the sets (t) and (d), which are common to the three algorithms, HILS-\emph{Complete} finds an average gap of $0.02\%$, compared to $0.29\%$ and $0.00\%$ for PH and DH-LDS, respectively.
For the sets (ta) and (ua), the difference of solution quality is also small, and both HILS-\emph{Complete} and DH-LDS return near-optimal solutions with average gaps below $0.15\%$.
For the (da) instances, DH-LDS produces solutions of higher quality ($1.17\%$ gap difference in average) than HILS-\emph{Complete}, albeit at the expense of a larger CPU time. HILS-\emph{Complete} attained 1819 out of the 1917 known optimal solutions, and a total of 9 new best upper bounds were also found for some open (da) instances.

\begin{figure}[htb]
	\centering
	\includegraphics[width=0.85\textwidth]{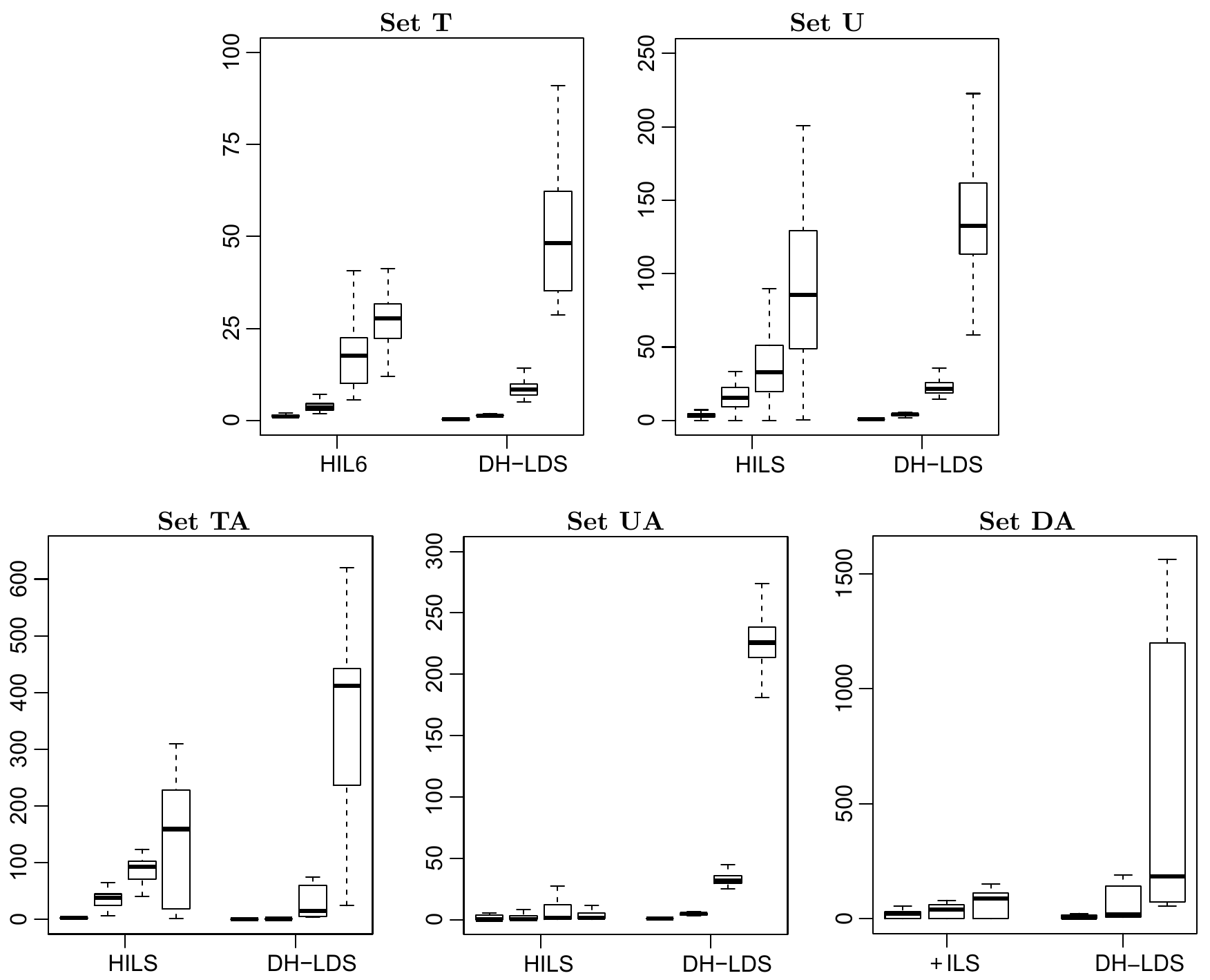}
	\caption{CPU time, in seconds, of HILS-\emph{Complete} and DH-LDS. One boxplot for each group $\times$ instance size in terms of number of items.}\label{fig:timeBox}
\end{figure}

DH-DLS performs a truncated search in the branch-and-price tree, which loses its optimality certificate but often keeps finding near-optimal solutions when the problem size allows for~it. Yet, as $n$ grows, its CPU grows very quickly as noted in \cite{Sadykov2013a}: ``a disadvantage of our primal heuristic is that the running time increases rapidly with the number of items''. 
To better evaluate the scalability of the methods, Figure~\ref{fig:timeBox} reports the CPU time of DH-DLS and HILS-\emph{Complete} for each instance set and size. Moreover, 
%The CPU time of DH-DLS is initially low for small instances, but tends to increase by a factor $10$ every time the instance size doubles. This effect is especially marked on the instance set (da), with a larger ratio of items per bin and arbitrary conflict graphs.
we fitted the CPU time as a power law $f(n) = \alpha \times n^\beta$ of the number of items $n$ for each instance class (least-squares regression of an affine function on the log-log graph). The observed CPU time of HILS-\emph{Complete} grows as $\cO(n^{1.49})$. In contrast, the time of DH-DLS grows at a rate which is faster than cubic for some instance sets, with power laws of the form $n^\beta$ where $2.45 \leq \beta \leq 3.62$.

\begin{figure}[htbp]
\centering
\includegraphics[width=0.88 \textwidth]{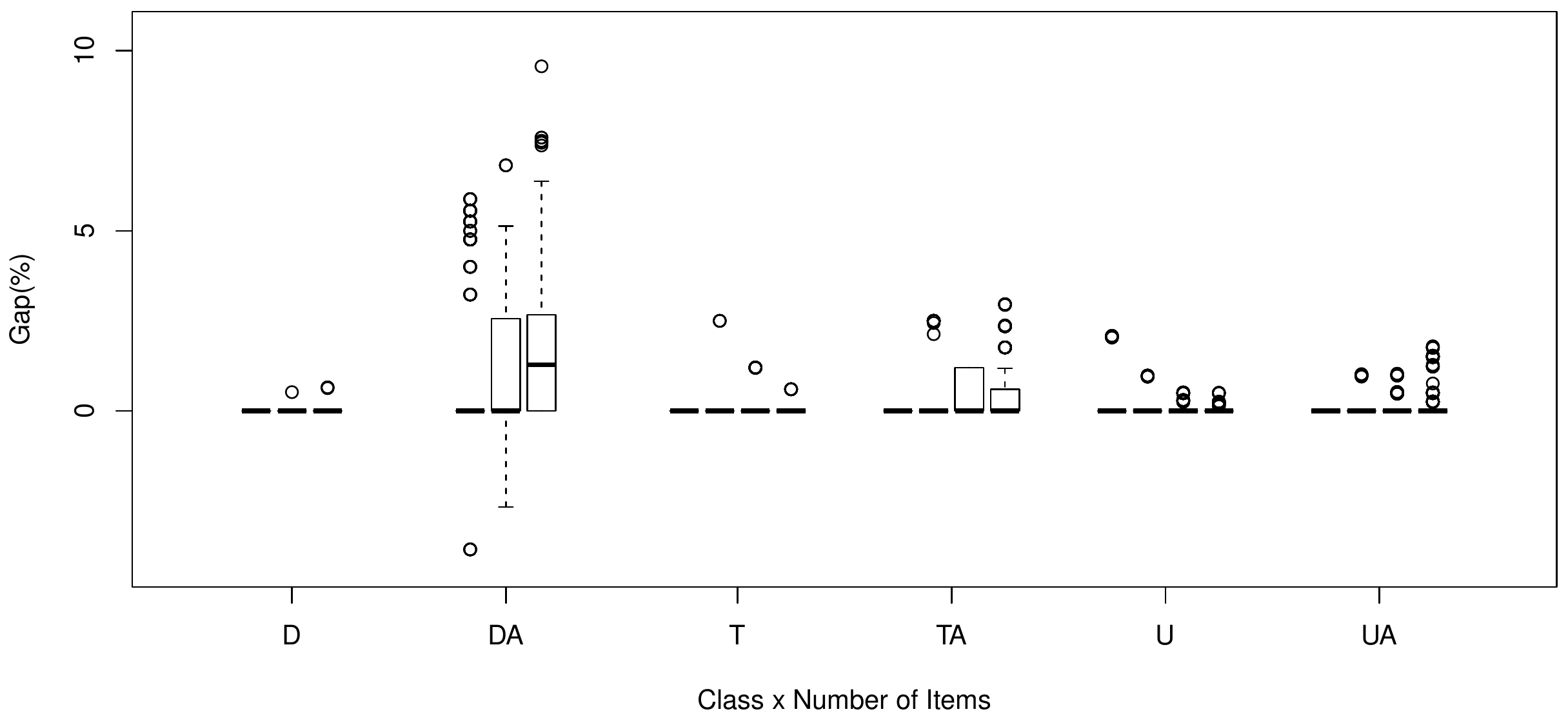}
\caption{HILS-\emph{Complete} GAP by each group of instance.}\label{fig:boxplot2}
\end{figure}

We also analyzed the impact of some instance characteristics on the performance of  HILS-\emph{Complete}.
Figure~\ref{fig:boxplot2} displays the boxplots of the percentage gap for each combination of class $\times$ number of items.
For the data sets (t), (u), and (d) with interval conflict graph, the optimal solution was attained by HILS-\emph{Complete} on nearly all instances. As a consequence, the boxplots are represented as a single line at a gap of $0.00\%$ and any other gap value appears as an outlier.
The other data sets, especially (ta) and (da) with arbitrary conflict graph, are more difficult, such that a gap of $0.00\%$ does not represent the average behavior of the method anymore. The difficulty of the instances tends to grow with their size, although this effect is counterbalanced by the way the ``gap'' metric is defined: on small instances with few bins in the optimal solution, a \emph{mistake} of one bin directly translates into a large gap (e.g., $5\%$ if the BKS includes 20 bins) leading to some outliers for small problems.

\begin{figure}[htbp]
	\centering
	\vspace*{0.15cm}
	\includegraphics[width= 0.88 \textwidth]{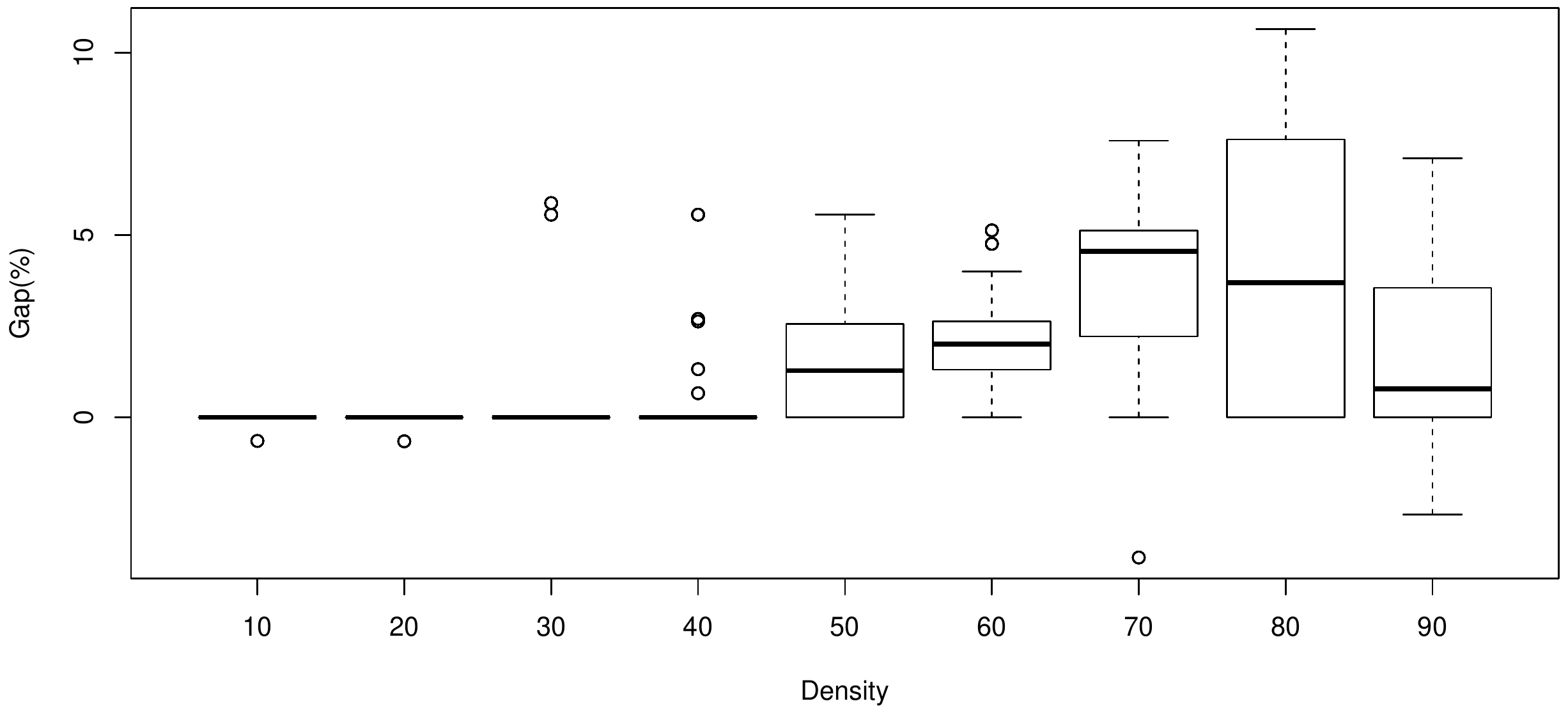}
	\caption{HILS-\emph{Complete} GAP by density of the conflict graph.} \label{fig:boxplot1}
\end{figure}

Instance class and size are not the only factors which impact the performance of the methods. The density of the conflict graph is also important. This effect is illustrated in Figure \ref{fig:boxplot1}, which displays the boxplots of the percentage gap as a function of density.
Generally, the most difficult instances are neither those with low density (equivalent to a bin packing problem), nor those with high density (equivalent to a tightly-constrained coloring problem), but those which combine the effects of both packing and conflict constraints at a density level between $50\%$ and $80\%$. HILS-\emph{Complete} appears to perform very well for problem instances with low density (up to $20\%$). On these instance, the BKS has been systematically reached and even sometimes improved, as highlighted by the two outliers located below $0.00\%$.

We also evaluated the impact of the large neighborhoods, comparing the results of ILS-\emph{Simple} and HILS-\emph{Complete}. We observe a significant difference of solution quality, confirmed by a pairwise Wilcoxon test (p-value $\leq 10^{-64}$). The large neighborhoods have a positive impact on solution quality, but consume some additional CPU time (49.85 seconds in average per instance with the large neighborhoods, and 12.53 seconds without). Note that we also conducted side experiments to investigate whether an increase in the number of iterations of ILS-\emph{Simple} could help to profit from this additional CPU time to reach better solutions, but the quality improvements were only minor when doubling the number of iterations \mbox{(see Section \ref{Analysis})}.

\begin{figure}[htbp]
	\centering
	\includegraphics[width= \textwidth]{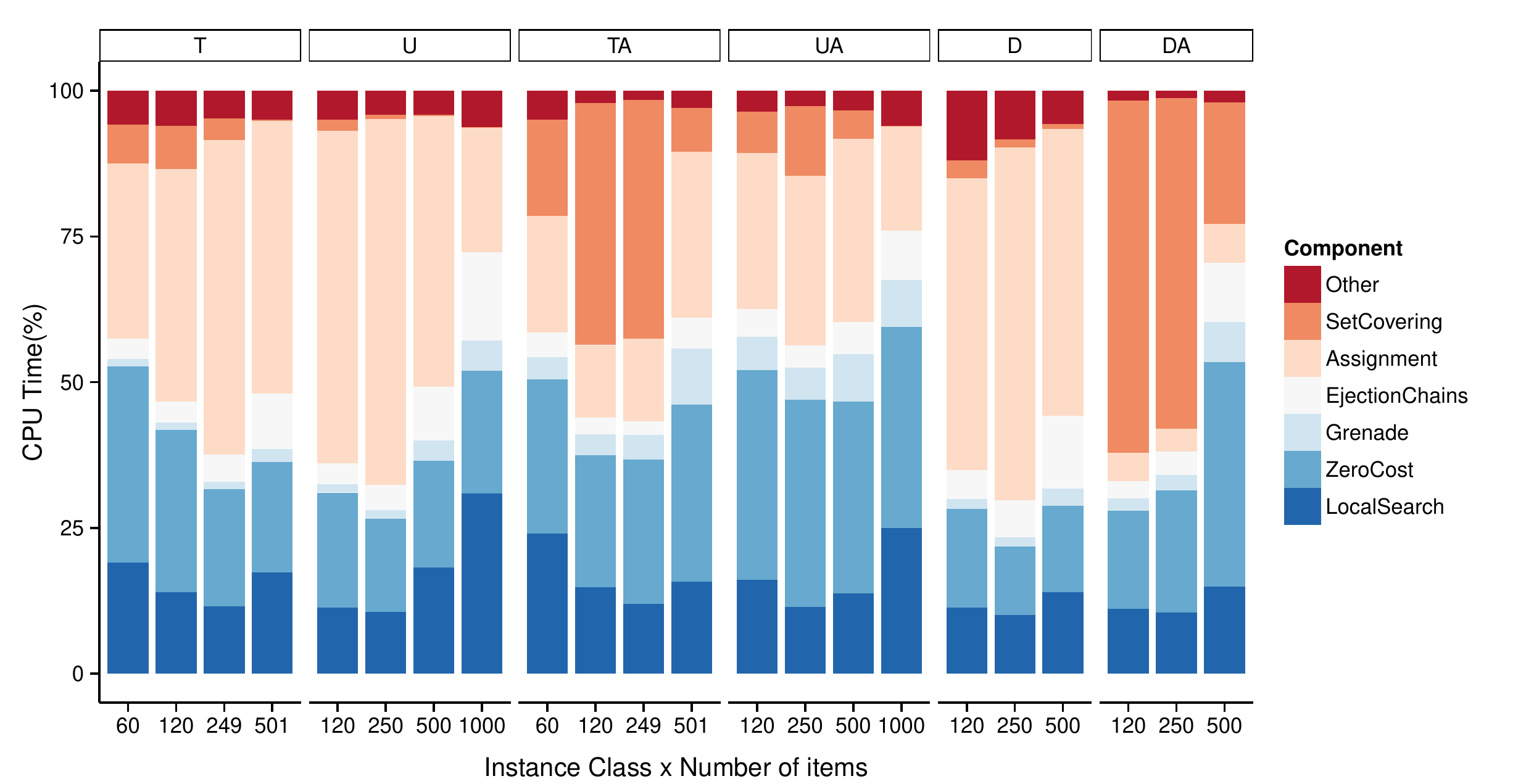}
	\caption{Time consumption of each component of HILS-\emph{Complete}.} \label{fig:barplot}
\end{figure}

Finally, we analyzed the time consumption of each component of the method.
The results of this analysis are reported in Figure \ref{fig:barplot}, which displays the percentage CPU time of each main component of the approach.  This metric has been aggregated per instance class $\times$ number of items. The printed article contains the figure in black and white, and we refer to the on-line article for a color figure. The search effort appears to be well distributed among the search components, i.e., no specific neighborhood consumes the majority of the time. Overall, the assignment neighborhood, local search and zero-cost moves are the largest time consumers. In contrast, the ejection chains and grenade neighborhoods use less than $6\%$ and $4\%$ of the CPU time, respectively.  The computational burden of the 0-cost moves mostly relates to the fact that it constitutes the first loop of the local search, and thus involves many moves evaluations which are later known as non-improving, and thus pruned in subsequent local search iterations. In line with the previous discussions, we also observe that the neighborhood based on set covering requires a larger amount of CPU time for difficult instances with an arbitrary conflict graph, such as~(ta)~and~(da).

Overall, from these experiments, HILS-\emph{Complete} is a promising alternative method for the BPPC, producing high quality solutions with an overall gap of $0.22\%$ from the BKS, and improving upon the previous best metaheuristic for the problem in a more controlled CPU time than existing mathematical programming-based methods. Having two alternative methods with their own strengths (near optimality for DH-LDS, and good scalability for HILS) will be a good asset for future works, which can explore new hybridizations. This option was kept aside in this work in order to better concentrate on the neighborhood aspects. Since these neighborhoods are central in the design of the method, we now go a step further to identify the most promising search components and measure the impact of its key parameters. This is the subject of a detailed analysis in the next~section.

\subsection{Sensitivity Analysis -- Search Parameters and Neighborhoods}
\label{Analysis}

This section analyses the sensibility of the method to a change of its parameters, and evaluates the contribution of its key components and neighborhoods. Starting from the standard HILS-\emph{Complete} configuration, we varied one parameter/factor at a time (OFAT approach) and tested the resulting configurations. These experiments are summarized in Tables \ref{tab:ShakingTermination}~to~\ref{tab:SetCovering}. Each line presents the performance of an alternative configuration of the method (indexed from~A~to~R), in terms of average gap and time, on the six classes of instances. To maintain a reasonable computational effort, a single run was performed for each configuration on the 2060 instances. \\

\textbf{Impact of the general search parameters.} Our first investigation concerns the shaking operator and the termination criterion, as these parameters are recurrent in all metaheuristics based on iterated local search. In the first two alternative configurations, we modified the periodicity of the shaking operator, i.e., the number $N_\textsc{ls}$ of zero-cost moves and local search phases before shaking. This parameter is halved in configuration A, and doubled in configuration~B. Since the product $N_\textsc{shak} \times N_\textsc{ls}$  governs the total number of local search descents before termination, we kept it constant  by updating $N_\textsc{shak}$ accordingly.
The configurations C and D aim to investigate the impact of longer and shorter runs, respectively, by doubling or halving the value $N_\textsc{shak}$ without impacting $N_\textsc{ls}$.
Finally, the configurations E, F and G investigate alternative strengths for the shaking operator, which may involve to perform stronger shaking for a better exploration of the search space ($S_\textsc{shak} = 5$), or weaker shaking for further intensification ($S_\textsc{shak}=1$ or $2$).

 \begin{table}[htbp]
	\renewcommand{\arraystretch}{1.25}
	\setlength{\tabcolsep}{0.039cm}
	\scalebox{0.82}
	{
		\begin{tabular}{|l|ccc|M{1cm}M{1cm}|M{1cm}M{1cm}|M{1cm}M{1cm}|M{1cm}M{1cm}|M{1cm}M{1cm}|M{1cm}c|} 
			\hline
			& \multirow{2}{*}{$N_\textsc{shak}$} & \multirow{2}{*}{$N_\textsc{ls}$} & \multirow{2}{*}{$S_\textsc{shak}$} & \multicolumn{2}{c|}{(t)} & \multicolumn{2}{c|}{(u)} & \multicolumn{2}{c|}{(ta)} & \multicolumn{2}{c|}{(ua)} & \multicolumn{2}{c|}{(d)} & \multicolumn{2}{c|}{(da)} \\ 
			&     &     &   & Gap            & Time   & Gap            & Time   & Gap            & Time   & Gap            & Time   & Gap            & Time   & Gap            & Time \\ \hline
			\rowcolor{gray!10} Standard & 50  & 100 & 3 & \textbf{0.02}  & 9.97   & 0.01           & 34.44  & 0.23           & 52.22  & 0.05           & 190.40 & \textbf{0.00}  & 11.78  & 1.16           & 31.45 \\ \hline
			A. More Frequent               & 100 & 50  & 3 & \textbf{0.02}  & 13.76  & 0.01           & 42.16  & 0.22           & 80.40  & 0.04           & 250.58 & \textbf{0.00}  & 13.81  & 1.16           & 57.97 \\
			B. Less Frequent               & 25  & 200 & 3 & 0.04           & 10.96  & 0.02           & 32.58  & 0.26           & 46.08  & 0.10           & 73.39  & \textbf{0.00}  & 13.05  & 1.40           & 22.70 \\
			C. Longer Run                  & 100 & 100 & 3 &\textbf{0.02}   & 21.03  & \textbf{0.00}  & 67.83  & 0.21           & 97.67  & \textbf{0.03}  & 382.53 & \textbf{0.00}  & 25.35  & \textbf{1.09}  & 59.95 \\
			D. Shorter Run                 & 25  & 100 & 3 & 0.04           & 5.65   & 0.02           & 20.18  & 0.28           & 34.69  & 0.10           & 50.05  & 0.01           & 6.45   & 1.47           & 15.60 \\
			E. Stronger Shaking            & 50  & 100 & 5 & 0.03           & 10.93  & 0.01           & 37.71  & 0.24           & 56.82  & 0.09           & 111.00 & \textbf{0.00}  & 12.87  & 1.17           & 31.18 \\
			F. Weaker Shaking 1            & 50  & 100 & 1 & \textbf{0.02}  & 10.56  & 0.01           & 31.57  & \textbf{0.18}  & 76.25  & 0.06           & 166.64 & \textbf{0.00}  & 12.47  & 1.42           & 25.78 \\
			G. Weaker Shaking 2            & 50  & 100 & 2 & 0.03           & 11.03  & 0.01           & 32.81  & 0.24           & 62.01  & 0.07           & 132.53 & \textbf{0.00}  & 12.73  & 1.29           & 30.31 \\ \hline
		\end{tabular}
	}
	\caption{Impact of some variations of the shaking and termination parameters.}
	\label{tab:ShakingTermination}
\end{table}

The changes of parameter setting for the configurations A, B, E, F and G are generally detrimental for the performance of the method. It appears that strengthening the shaking, or calling upon this operator more frequently has only a small impact on performance, but the inverse process would have a larger negative impact. The configuration D allows a longer run, and thus naturally leads to better solutions. This improvement of solution quality, however, remains minor in comparison to the additional CPU time. \\

\textbf{Contribution of the large neighborhoods.} The second experiment aims to evaluate the respective role of each large neighborhood as well as the \emph{0-cost} moves, which are critical for search diversification. We thus deactivated, in the configurations H to K, each one of the four large neighborhoods. In configuration L, we deactivated all four neighborhoods. Finally, no 0-cost moves are applied in configuration N.

\begin{table}[htbp]
	\renewcommand{\arraystretch}{1.25}
	\setlength{\tabcolsep}{0.05cm}
	\scalebox{0.84}
	{
		\begin{tabular}{|l|M{1cm}M{1cm}|M{1cm}M{1cm}|M{1cm}M{1cm}|M{1cm}M{1cm}|M{1cm}M{1cm}|M{1cm}c|} 
			\hline
			& \multicolumn{2}{c|}{(t)} & \multicolumn{2}{c|}{(u)} & \multicolumn{2}{c|}{(ta)} & \multicolumn{2}{c|}{(ua)} & \multicolumn{2}{c|}{(d)} &\multicolumn{2}{c|}{(da)} \\ 
			& Gap           & Time  & Gap            & Time  & Gap           & Time  & Gap           & Time   & Gap           & Time  & Gap           & Time \\ \hline
			\rowcolor{gray!10} Standard              & \textbf{0.02}  & 9.97  & \textbf{0.01}  & 34.44  & \textbf{0.23}  & 52.22 & \textbf{0.05}  & 190.40 & \textbf{0.00}  & 11.78  & \textbf{1.16}  & 31.45 \\ \hline              
			H. No Assign    &  0.03           &  5.21  & \textbf{0.01}  & 24.38 & 0.27          & 43.48 & 0.10          & 61.89  & \textbf{0.00} & 4.73  & 1.29           &  28.37 \\
			I. No Ejection Chains  & \textbf{0.02} & 9.99  & \textbf{0.01}  & 30.92 & 0.24 & 56.67 & 0.07          & 108.31 & \textbf{0.00} & 11.58 & 1.23          & 28.87 \\
			J. No Grenade   & \textbf{0.02}          & 9.84  & 0.02           & 34.99 & 0.25          & 46.49 & 0.07          & 70.97  & 0.01          & 11.58 & 1.21 & 29.51 \\
			K. No Set Covering  & 0.37  & 9.28   & 0.03  & 33.16  & 0.54  & 33.76  & 0.30  & 46.34 & \textbf{0.00}  & 11.88  & 1.40  & 13.75 \\   
			L. No Large Neighborhoods & 0.49          & 4.50  & 0.24           & 18.66 & 0.72          & 11.35 & 0.43          & 18.98  & 0.07          & 3.97  & 1.36          & 7.39 \\
			M. No Large Neighborhoods ++ & 0.39 & 50.21 & 0.15 & 65.62 & 0.65 & 27.61 & 0.39 & 47.52 & 0.05 & 15.56 & 1.22 & 16.33  \\
			N. No 0-cost Moves        & 0.10          & 9.52  & 0.02           & 32.63 & 0.31          & 57.97 & 0.07          & 172.05 & \textbf{0.00} & 10.15 & 3.16          & 17.06 \\
			\hline
		\end{tabular}
	}
	\caption{Sensitivity analysis when deactivating some neighborhoods.}
	\label{tab:Neighboor}
\end{table}

All search components appear to contribute positively to the final solution quality.
By order of magnitude, the largest loss occurs when deactivating the 0-cost moves, as they contribute very significantly to diversity the search. In second position, deactivating all large neighborhoods (as in ILS-\emph{Simple}) also leads to a large loss of solution quality, but also to a gain of computational time. We tested longer runs (configuration M, via a fourfold increase of $N_\textsc{shak}$) without the large neighborhoods, and this was not sufficient to regain the solution quality of the complete method.

The large neighborhood based on set covering has the largest impact on the search, especially for the instance classes (t) and (ua). As noted in Figure \ref{fig:barplot}, on these particular instances, the 
set covering problem is conveniently solved in small CPU time. This allows to use a larger pool of columns, thus increasing the chances of solution improvement.
Finally, the last three neighborhoods, ejection chains, assignment, and grenade have a more moderate impact on solution quality. Their contributions are only observed on the difficult instances: (ta), (ua) and (da). Nevertheless, note that the removal of these three neighborhoods together is more largely detrimental (compare configurations K and L). The three neighborhoods provide different forms of solutions improvement and diversification, which are not critical when taken one by one, but much more significant when considered as a whole. \\

\textbf{Impact of the set-covering parameters.} The last parameters of the method concern the set covering neighborhood. Our preliminary calibration led to a time limit $T_\textsc{Limit} = 20$ seconds for the solver, as well as an adaptive mechanism for the size of the pool, starting from a pool size of $S_\textsc{pool} = 1500$ columns. We thus tested the impact of this adaptive mechanism in the configurations O to Q, by evaluating three static pool size values of 1000, 1500 and 2000 columns, along with a time of 10, 20 and 60 seconds, respectively, for the resolution of the set covering problem.
Finally, the last configuration R considers a possible replacement of the set covering by a set partitioning formulation, using an equality in Equation (\ref{eq:r1SP}).

\begin{table}[htbp]
	\renewcommand{\arraystretch}{1.25}
	\setlength{\tabcolsep}{0.04cm}
	\scalebox{0.82}
	{
		\begin{tabular}{|l|cHcc|M{1cm}M{1cm}|M{1cm}M{1cm}|M{1cm}M{1cm}|M{1cm}M{1cm}|M{1cm}M{1cm}|M{1cm}c|} 
			\hline
			& \multirow{2}{*}{Adapt} & \multirow{2}{*}{$N_\textsc{sc}$} & \multirow{2}{*}{$S_\textsc{pool}$} & \multirow{2}{*}{$T_\textsc{Limit}$} & \multicolumn{2}{c|}{(t)} & \multicolumn{2}{c|}{(u)} & \multicolumn{2}{c|}{(ta)} & \multicolumn{2}{c|}{(ua)} & \multicolumn{2}{c|}{(d)} & \multicolumn{2}{c|}{(da)} \\ 
			& & & & & Gap  & Time & Gap   & Time & Gap   & Time &  Gap  & Time & Gap   & Time & Gap   & Time \\ \hline
			\rowcolor{gray!10}  Standard       & Yes & 25 & 1500 & 20 & \textbf{0.02}  & 9.97  & \textbf{0.01}  & 34.44  & 0.23  & 52.22 & \textbf{0.05}  & 190.40 & \textbf{0.00}  & 11.78  & \textbf{1.16}  & 31.45 \\ \hline
			O. Static Pool --       & No & 25 & 1000 & 10 & 0.05  & 9.80   & 0.02  & 34.14  & 0.33  & 41.57  & 0.16  & 44.97 & \textbf{0.00}  & 11.84  & 1.30  & 16.58 \\
			P. Static    & No & 25 & 1500 & 20 & 0.04  & 9.85   & 0.02  & 35.18  & 0.26  & 54.75  & 0.14  & 48.92 & \textbf{0.00}  & 11.80  & 1.20  & 24.84 \\
			Q. Static Pool +  & No & 25 & 2000 & 60 & 0.03  & 11.42  & 0.02  & 34.29  & \textbf{0.20}  & 95.66  & 0.11 & 71.90 & \textbf{0.00}  & 11.93  & 1.28  & 58.21 \\
			R. Set Partitioning & Yes & 25 & 1500 & 20 & \textbf{0.02}  & 10.19  & \textbf{0.01}  & 33.85  & 0.22  & 52.82  & 0.07  & 122.83 & \textbf{0.00}  & 12.13  & 1.20  & 27.71 \\
			\hline
		\end{tabular}
	}
	\caption{Sensitivity analysis on the parameters of the set covering neighborhood.}
	\label{tab:SetCovering}
\end{table}

In these experiments, we observe that the size of the pool is a very sensible parameter.
A small value leads to simple set covering problems which are not likely to result in solution improvements, even when solved to optimality, while a larger value opens the way to more improvement opportunities, at the risk of not solving the mathematical model in the allowed time. For the instances of class (da), an intermediate pool size of 1500 appears to be desirable, while for other instances, e.g., (ta) and (ua), a larger pool size with 2000 columns leads to solutions of better quality at the cost of additional computational time. Clearly, the best parameter setting is dependent on the instance characteristics. This led us to use the adaptive mechanism, which performs better overall than each of the static configurations. Finally, using either the set covering or the set partitioning model did not lead to a significant difference of solution quality.

\section{Conclusions}
\label{sec:conclusion}

In this paper, we introduced an ILS based on several classes of local and large (exponential-sized) neighborhoods for the BPPC. We intentionally relied on simple metaheuristic concepts so as to draw the complete focus on the large neighborhoods and accurately measure their contribution to the search.
We proposed $\cO(1)$ move evaluation procedures for the local search, polynomial variants of ejection chains and assignment neighborhoods, as well as an adaptive search mechanism for the set covering-based neighborhood. We also introduced a controlled use of 0-cost moves to further~diversify~the~search.

The resulting ILS produces solution of good quality on the classic BPPC benchmark instances, with an average gap to the BKS of $0.22\%$. This is a better solution quality than the previous best metaheuristic, the genetic algorithm of \cite{Muritiba2009}, and a slightly lower quality than the diving heuristic with limited discrepancy search (DH-LDS) of \cite{Sadykov2013a}. Still, the current instances are an ideal testing ground for column-generation-based methods, e.g. DH-LDH, as they contain very few items per bin (2.87 in average). On the other hand, the proposed metaheuristic exploits different search concepts, and scales very well with an increase of problem size, with a measured CPU time in $\cO(n^{1.49})$. In comparison, the CPU time of DH-LDH tends to increase by a factor ten for each twofold increase of problem size. Finally, we conducted extensive experiments to measure the contribution of each neighborhood as well as the impact of the key search parameters. As underlined by these experiments, the 0-cost moves are critical for the search performance, as well as the set covering-based neighborhood. The other large neighborhoods, i.e., ejection chains, assignment, and grenade, have a smaller individual impact but contribute significantly to the search when used together. 

Many research perspectives are open on this class of problems.
First, future hybridizations should allow to combine the strengths of both neighborhood-centered search and column-generation based-heuristics, with the aim of keeping both near-optimality and tractability. Future research should thus be done along the path of \emph{matheuristics}, aiming to harness the strength of mathematical programming and neighborhood search. Second, populations of solutions and more advanced memory structures can be a strong asset to help diversifying the search and favoring the discovery of better solutions. Finally, as the current benchmark instances are now solved to near-optimality, new and larger test sets, in terms of overall number of items, and number of items per bin, should also be investigated.

%\bibliographystyle{ormsv080-noURLDOI}
%\bibliography{references}{}

\begin{thebibliography}{41}
\expandafter\ifx\csname natexlab\endcsname\relax\def\natexlab#1{#1}\fi
\expandafter\ifx\csname url\endcsname\relax
  \def\url#1{{\tt #1}}\fi
\expandafter\ifx\csname urlprefix\endcsname\relax\def\urlprefix{URL }\fi
\expandafter\ifx\csname urlstyle\endcsname\relax
  \expandafter\ifx\csname doi\endcsname\relax
  \def\doi#1{doi:\discretionary{}{}{}#1}\fi \else
  \expandafter\ifx\csname doi\endcsname\relax
  \def\doi{doi:\discretionary{}{}{}\begingroup \urlstyle{rm}\Url}\fi \fi

\bibitem[{Ahuja et~al.(2002)Ahuja, Ergun, Orlin, and Punnen}]{Ahuja2002a}
Ahuja, R.~K., O.~Ergun, J.~B. Orlin, A.~P. Punnen. 2002.
\newblock A survey of very large-scale neighborhood search techniques.
\newblock {\it Discrete Applied Mathematics\/} {\bf 123}(1-3) 75--102.

\bibitem[{Avanthay et~al.(2003)Avanthay, Hertz, and Zufferey}]{Avanthay2003}
Avanthay, C., A.~Hertz, N.~Zufferey. 2003.
\newblock A variable neighborhood search for graph coloring.
\newblock {\it European Journal of Operational Research\/} {\bf 151}(2)
  379--388.

\bibitem[{Bl{\"{o}}chliger and Zufferey(2008)}]{Blochliger2008}
Bl{\"{o}}chliger, I., N.~Zufferey. 2008.
\newblock A graph coloring heuristic using partial solutions and a reactive
  tabu scheme.
\newblock {\it Computers {\&} Operations Research\/} {\bf 35}(3) 960--975.

\bibitem[{Blum and Roli(2003)}]{Blum2003}
Blum, C., A.~Roli. 2003.
\newblock Metaheuristics in combinatorial optimization: overview and conceptual
  comparison.
\newblock {\it ACM Computing Surveys\/} {\bf 35}(3) 268--308.

\bibitem[{Deineko and Woeginger(2000)}]{Deineko2000}
Deineko, V.~G., G.~J. Woeginger. 2000.
\newblock A study of exponential neighborhoods for the travelling salesman
  problem and for the quadratic assignment problem.
\newblock {\it Mathematical Programming\/} {\bf 87}(3) 519--542.

\bibitem[{Delorme et~al.(2016)Delorme, Iori, and Martello}]{Delorme2016}
Delorme, Maxence, Manuel Iori, Silvano Martello. 2016.
\newblock {Bin packing and cutting stock problems: Mathematical models and
  exact algorithms}.
\newblock {\it European Journal of Operational Research\/} {\bf 255}(1) 1--20.

\bibitem[{Dowsland and Thompson(2008)}]{Dowsland2008}
Dowsland, K.A., J.M. Thompson. 2008.
\newblock An improved ant colony optimisation heuristic for graph colouring.
\newblock {\it Discrete Applied Mathematics\/} {\bf 156}(3) 313--324.

\bibitem[{Elhedhli et~al.(2011)Elhedhli, Li, Gzara, and
  Naoum-Sawaya}]{Elhedhli2011}
Elhedhli, S., L.~Li, M.~Gzara, J.~Naoum-Sawaya. 2011.
\newblock A branch-and-price algorithm for the bin packing problem with
  conflicts.
\newblock {\it INFORMS Journal on Computing\/} {\bf 23}(3) 404--415.

\bibitem[{Epstein and Levin(2008)}]{Epstein2008a}
Epstein, L., A.~Levin. 2008.
\newblock On bin packing with conflicts.
\newblock {\it SIAM Journal on Optimization\/} {\bf 19}(3) 1270--1298.

\bibitem[{Epstein et~al.(2008)Epstein, Levin, and Stee}]{Epstein2008}
Epstein, L., A.~Levin, R.~Stee. 2008.
\newblock Two-dimensional packing with conflicts.
\newblock {\it Acta Informatica\/} {\bf 45}(3) 155--175.

\bibitem[{Falkenauer(1996)}]{Falkenauer1996}
Falkenauer, E. 1996.
\newblock A hybrid grouping genetic algorithm for bin packing.
\newblock {\it Journal of Heuristics\/} {\bf 2}(1) 5--30.

\bibitem[{{Fernandes Muritiba} et~al.(2010){Fernandes Muritiba}, Iori,
  Malaguti, and Toth}]{Muritiba2009}
{Fernandes Muritiba}, A.E., M.~Iori, E.~Malaguti, P.~Toth. 2010.
\newblock {Algorithms for the bin packing problem with conflicts}.
\newblock {\it INFORMS Journal on Computing\/} {\bf 22}(3) 401--415.

\bibitem[{Fleszar and Charalambous(2011)}]{Fleszar2011}
Fleszar, K., C.~Charalambous. 2011.
\newblock Average-weight-controlled bin-oriented heuristics for the
  one-dimensional bin-packing problem.
\newblock {\it European Journal of Operational Research\/} {\bf 210}(2)
  176--184.

\bibitem[{Galinier and Hao(1999)}]{Galinier1999}
Galinier, P., J.~Hao. 1999.
\newblock Hybrid evolutionary algorithms for graph coloring.
\newblock {\it Journal of combinatorial optimization\/} {\bf 3}(4) 379--397.

\bibitem[{Gendreau et~al.(2004)Gendreau, Laporte, and Semet}]{Gendreau2004}
Gendreau, M., G.~Laporte, F.~Semet. 2004.
\newblock Heuristics and lower bounds for the bin packing problem with
  conflicts.
\newblock {\it Computers \& Operations Research\/} {\bf 31}(3) 347--358.

\bibitem[{Glover(1996)}]{Glover1996a}
Glover, F. 1996.
\newblock Ejection chains, reference structures and alternating path methods
  for traveling salesman problems.
\newblock {\it Discrete Applied Mathematics\/} {\bf 65}(1-3) 223--253.

\bibitem[{Gutin(1999)}]{Gutin1999}
Gutin, G. 1999.
\newblock Exponential neighbourhood local search for the traveling salesman
  problem.
\newblock {\it Computers {\&} Operations Research\/} {\bf 26}(4) 313--320.

\bibitem[{Hamdi-Dhaoui et~al.(2014)Hamdi-Dhaoui, Labadie, and
  Yalaoui}]{Hamdi-Dhaoui2014}
Hamdi-Dhaoui, K., N.~Labadie, A.~Yalaoui. 2014.
\newblock The bi-objective two-dimensional loading vehicle routing problem with
  partial conflicts.
\newblock {\it International Journal of Production Research\/} {\bf 52}(19)
  5565--5582.

\bibitem[{Hertz and de~Werra(1987)}]{Hertz1987}
Hertz, A., D.~de~Werra. 1987.
\newblock Using tabu search techniques for graph coloring.
\newblock {\it Computing\/} {\bf 39}(4) 345--351.

\bibitem[{Jansen(1999)}]{Jansen1999}
Jansen, K. 1999.
\newblock An approximation scheme for bin packing with conflicts.
\newblock {\it Journal of Combinatorial Optimization\/} {\bf 3}(4) 363--377.

\bibitem[{Jansen and {\"{O}}hring(1997)}]{Jansen1997}
Jansen, K., S.~{\"{O}}hring. 1997.
\newblock Approximation algorithms for time constrained scheduling.
\newblock {\it Information and Computation\/} {\bf 132}(2) 85--108.

\bibitem[{Khanafer et~al.(2010)Khanafer, Clautiaux, and Talbi}]{Khanafer2010}
Khanafer, A., F.~Clautiaux, E.~Talbi. 2010.
\newblock New lower bounds for bin packing problems with conflicts.
\newblock {\it European Journal of Operational Research\/} {\bf 206}(2)
  281--288.

\bibitem[{Kuhn(1955)}]{kuhn1955}
Kuhn, H.~W. 1955.
\newblock The hungarian method for the assignment problem.
\newblock {\it Naval research logistics quarterly\/} {\bf 2}(1-2) 83--97.

\bibitem[{Laporte and Desroches(1984)}]{Laporte1984}
Laporte, G., S.~Desroches. 1984.
\newblock Examination timetabling by computer.
\newblock {\it Computers \& Operations Research\/} {\bf 11}(4) 351--360.

\bibitem[{Lewis et~al.(2012)Lewis, Thompson, Mumford, and Gillard}]{Lewis2012}
Lewis, R., J.~Thompson, C.~Mumford, J.~Gillard. 2012.
\newblock A wide-ranging computational comparison of high-performance graph
  colouring algorithms.
\newblock {\it Computers {\&} Operations Research\/} {\bf 39}(9) 1933--1950.

\bibitem[{Lewis(2016)}]{Lewis2016}
Lewis, R.M.R. 2016.
\newblock {\it A guide to graph colouring: algorithms and applications\/}.
\newblock Springer International Publishing.

\bibitem[{Malaguti et~al.(2008)Malaguti, Monaci, and Toth}]{Malaguti2008}
Malaguti, E., M.~Monaci, P.~Toth. 2008.
\newblock A metaheuristic approach for the vertex coloring problem.
\newblock {\it INFORMS Journal on Computing\/} {\bf 20}(2) 302--316.

\bibitem[{Malaguti and Toth(2010)}]{Malaguti2010}
Malaguti, E., P.~Toth. 2010.
\newblock A survey on vertex coloring problems.
\newblock {\it International Transactions in Operational Research\/} {\bf
  17}(1) 1--34.

\bibitem[{Masson et~al.(2013)Masson, Vidal, Michallet, Penna, Petrucci,
  Subramanian, and Dubedout}]{Masson2012a}
Masson, R., T.~Vidal, J.~Michallet, P.~H.~V. Penna, V.~Petrucci,
  A.~Subramanian, H.~Dubedout. 2013.
\newblock An iterated local search heuristic for multi-capacity bin packing and
  machine reassignment problems.
\newblock {\it Expert Systems with Applications\/} {\bf 40}(13) 5266--5275.

\bibitem[{Minh et~al.(2013)Minh, Hoai, and Nguyet}]{Minh2013}
Minh, T.~T., T.~V. Hoai, T.~T.~N. Nguyet. 2013.
\newblock A memetic algorithm for waste collection vehicle routing problem with
  time windows and conflicts.
\newblock B.~Murgante, ed., {\it Computational Science and Its Applications --
  ICCSA 2013\/}. Springer, Berlin, Heidelberg, 485--499.

\bibitem[{Monaci and Toth(2006)}]{Monaci2006}
Monaci, M., P.~Toth. 2006.
\newblock {A set-covering-based heuristic approach for bin-packing problems}.
\newblock {\it INFORMS Journal on Computing\/} {\bf 18}(1) 71--85.

\bibitem[{Morgenstern(1996)}]{Morgenstern1996}
Morgenstern, C. 1996.
\newblock Distributed coloration neighborhood search.
\newblock D.S. Johnson, M.A. Trick, eds., {\it Discrete Mathematics and
  Theoretical Computer Science\/}. American Mathematical Society, Providence,
  RI, 335--358.

\bibitem[{Muter et~al.(2010)Muter, Birbil, and Sahin}]{Muter2010}
Muter, I., S.~I. Birbil, G.~Sahin. 2010.
\newblock {Combination of metaheuristic and exact algorithms for solving set
  covering-type optimization problems}.
\newblock {\it INFORMS Journal on Computing\/} {\bf 22}(4) 603--619.

\bibitem[{Quiroz-Castellanos et~al.(2015)Quiroz-Castellanos, Cruz-Reyes,
  Torres-Jimenez, G\'{o}mez, Huacuja, and Alvim}]{Quiroz-Castellanos2015}
Quiroz-Castellanos, M., L.~Cruz-Reyes, J.~Torres-Jimenez, C.~G\'{o}mez,
  H.~J.~F. Huacuja, A.~C.F. Alvim. 2015.
\newblock A grouping genetic algorithm with controlled gene transmission for
  the bin packing problem.
\newblock {\it Computers \& Operations Research\/} {\bf 55} 52--64.

\bibitem[{Sadykov and Vanderbeck(2013)}]{Sadykov2013a}
Sadykov, R., F.~Vanderbeck. 2013.
\newblock Bin packing with conflicts: a generic branch-and-price algorithm.
\newblock {\it INFORMS Journal on Computing\/} {\bf 25}(2) 244--255.

\bibitem[{Sarvanov and Doroshko(1981)}]{Sarvanov1981}
Sarvanov, V.I., N.N. Doroshko. 1981.
\newblock {The approximate solution of the travelling salesman problem by a
  local algorithm with scanning neighborhoods of factorial cardinality in cubic
  time (in Russian)}.
\newblock {\it Software: Algorithms and Programs 31\/}. Mathematical Institute
  of the Belarusian Academy of Sciences, Minsk, 11--13.

\bibitem[{S{\"{o}}rensen(2015)}]{Sorensen2015}
S{\"{o}}rensen, K. 2015.
\newblock Metaheuristics -- the metaphor exposed.
\newblock {\it International Transactions in Operational Research\/} {\bf
  22}(1) 3--18.

\bibitem[{Subramanian et~al.(2013)Subramanian, Uchoa, and
  Ochi}]{Subramanian2013a}
Subramanian, A., E.~Uchoa, L.S. Ochi. 2013.
\newblock {A hybrid algorithm for a class of vehicle routing problems}.
\newblock {\it Computers {\&} Operations Research\/} {\bf 40}(10) 2519--2531.

\bibitem[{Thompson and Psaraftis(1993)}]{Thompson1993}
Thompson, P.M., H.N. Psaraftis. 1993.
\newblock Cyclic transfer algorithms for multi-vehicle routing and scheduling
  problems.
\newblock {\it Operations Research\/} {\bf 41}(5) 935--946.

\bibitem[{Toth and Tramontani(2008)}]{TothTramontani2008}
Toth, P., A.~Tramontani. 2008.
\newblock An integer linear programming local search for capacitated vehicle
  routing problems.
\newblock B.~Golden, S.~Raghavan, E.~Wasil, eds., {\it The vehicle routing
  problem: latest advances and new challenges\/}. Springer, Boston, MA,
  275--295.

\bibitem[{Vidal et~al.(2013)Vidal, Crainic, Gendreau, and Prins}]{Vidal2012a}
Vidal, T., T.G. Crainic, M.~Gendreau, C.~Prins. 2013.
\newblock Heuristics for multi-attribute vehicle routing problems: a survey and
  synthesis.
\newblock {\it European Journal of Operational Research\/} {\bf 231}(1) 1--21.

\end{thebibliography}

\end{document}